\documentclass[11pt, oneside]{article}   	% use "amsart" instead of "article" for AMSLaTeX format
\usepackage{geometry}                		% See geometry.pdf to learn the layout options. There are lots.
\usepackage{graphicx}				% Use pdf, png, jpg, or eps§ with pdflatex; use eps in DVI mode
\usepackage{amssymb}
\usepackage{amsmath}
\usepackage{hyperref}
\usepackage{tabularx,booktabs}
\usepackage{color}
\usepackage{authblk}

\title{Adding Constraints to Bayesian Inverse Problems}
\author[1]{Jiacheng Wu}
\author[2, 3]{Jian-Xun Wang}
\author[1]{Shawn C. Shadden}
\affil[1]{Department of Mechanical Engineering, University of California, Berkeley}
\affil[2]{Department of Aerospace and Mechanical Engineering, University of Notre Dame}
\affil[3]{Center of Informatics and Computational Science, University of Notre Dame}

\date{}

\begin{document}
\maketitle
%\section{}
%\subsection{}

\section*{Abstract}
%\begin{abstract}
Using observation data to estimate unknown parameters in computational models is broadly important. This task is often challenging because solutions are non-unique due to the complexity of the model and limited observation data. However, the parameters or states of the model are often known to satisfy additional constraints beyond the model. Thus, we propose an approach to improve parameter estimation in such inverse problems by incorporating constraints in a Bayesian inference framework. Constraints are imposed by constructing a likelihood function based on fitness of the solution to the constraints. The posterior distribution of the parameters conditioned on (1) the observed data and (2) satisfaction of the constraints is obtained, and the estimate of the parameters is given by the maximum a posteriori estimation or posterior mean. Both equality and inequality constraints can be considered by this framework, and the strictness of the constraints can be controlled by constraint uncertainty denoting a confidence on its correctness. Furthermore, we extend this framework to an approximate Bayesian inference framework in terms of the ensemble Kalman filter method, where the constraint is imposed by re-weighing the ensemble members based on the likelihood function. A synthetic model is presented to demonstrate the effectiveness of the proposed method and in both the exact Bayesian inference and ensemble Kalman filter scenarios, numerical simulations show that imposing constraints using the method presented improves identification of the true parameter solution among multiple local minima. 
%\end{abstract}

%This framework of imposing constraint is implemented with respect to a nonlinear test problem with multiple local minima. 
%In both the exact Bayesian inference and ensemble Kalman filter scenarios, the results shows that imposing constraint helps to pick out the true parameter value among multiple local minima. We also show how the variances of the constraint constraint influences the overall convergence.

\section{Introduction}
%{\color{blue}Outline:\\
%\begin{itemize}
%	\item Bayesian inference framework has been widely used for parameter estimation: list some application papers.
%	\item Exact Bayesian with MCMC sampling (expensive)
%	\item Approximate Bayesian: EnKF / Sequential MC
%	\item However, the common issue is ill-posed. -- non-uniqueness, insufficient observation data..
%	\item Solution: additional information needs to be incorporated. Constraint, regularization. 		
%	\item In physical modeling, physics-based constraint exist.
%	\item How to incorporate? Some people try, previous work limited E.p, Linear constraint in Kalman filter. Why limited: Very difficult to incorporate Eq which is very complex, derivative is very hard to obtain, sometime the constraint is not hard constraint. 
%	\item In this paper, we propose a method that can naturally consider the constraints within the Bayesian framework. 
%	\item It can handle many types of constraint. Derivative-free. It can consider the UQ in the constraint.
%\end{itemize}
%}
Computational models are pervasively used in wide-ranging engineering applications. Recent advances in computer platforms and numerical methods have enabled models to be increasingly sophisticated and comprehensive.%~\cite{witherden2017future, morris2016computational, blocken201450}. 
With greater model complexity comes greater challenge to determine model parameters (including initial/boundary conditions), which are often unknown or uncertain. To address this challenge, one typically solves an {\em inverse problem} by using observational data to specify model parameters so that the model output matches the observational data. Many inversion techniques have been developed and used by different communities. These methods can be roughly categorized into two classes: variational and statistical approaches~\cite{asch2016data}. The variational approach aims to minimize a specific cost function based on classical optimization theory and calculus of variations~\cite{smedstad1991variational}, while the statistical approach aims to evaluate or maximize posterior functions based on statistics and Bayesian theory~\cite{cotter2009bayesian}. 

Because of its robustness and capability for uncertainty quantification, Bayesian inversion techniques are widely used for hidden state and parameter estimation for many physical systems~\cite{iglesias2014well,wang2017inferring,wang2016data,li2017data}. In the Bayesian framework, both the hidden state/parameters (prior) and observable quantities (likelihood) are described as random variables with statistical distributions. The Bayesian estimation aims to calculate the posterior distributions of the inferred quantities from the prior and likelihood based on Bayes theorem. Directly computing the posterior distribution based on the prior and likelihood functions is referred to as the exact Bayesian approach. In general, the posterior is obtained by sampling the prior and likelihood distributions based on efficient Monte Carlo sampling such as the Markov chain Monte Carlo (MCMC) method. However, since MCMC requires an enormous amount of samples, which is computationally infeasible when the likelihood calculation involves expensive model evaluations, many approximate Bayesian inversion approaches have been developed, such as the extended Kalman filter (EKF)~\cite{haykin2004kalman}, unscented Kalman filter (UKF)~\cite{wan2000unscented}, ensemble Kalman filter (EnKF)~\cite{evensen2003ensemble}, and sequential Monte Carlo (SMC) method~\cite{del2006sequential}.

A common challenge in solving inverse problems is identifiability. Measurement data is typically very limited and solutions for the hidden states and parameters are nonunique. Moreover, numerical stability of the inversion can be significantly reduced for ill-posed problems and uncontrolled inference may happen due to small random noise in the observation data. To address these issues, a general strategy is to incorporate additional information into the inversion process, either by including more observation data or imposing additional constraints. In most mission-critical applications, data are difficult to collect and limited in quality. In such cases, additional constraints can be significantly useful to help regularize the inversion results to consistent ranges and relieve ill-posedness. Fortunately, for many physical systems, constraints on the state and parameters are available based on existed knowledge~\cite{he2016numerical}. Nonetheless, most existing Bayesian methods do not take constraints into account~\cite{shao2010constrained}. 

Initial progress has been made to incorporate constraints into certain Bayesian filters. For example, Simon et al.~\cite{simon2002kalman} considered equality constraints in the standard Kalman filter by projecting the Kalman updated solution onto the state constraint surface. Shao et al.~\cite{shao2010constrained} developed a constrained sequential Monte Carlo algorithm based on acceptance/rejection and an optimization strategy. Most recently, Gardner et al.~\cite{gardner2014bayesian} also considered inequality constraints in the context of Bayesian optimization. However, the existing approaches to incorporate constraints have been developed for each specific Bayesian filter, and most of them are based on a linearized form of the constraints, which is limiting when constraint functions are complicated and highly nonlinear. Moreover, in many complex systems, the constraints are approximations to reality and formulating the constraint in a deterministic way may neglect the uncertainties associated with constraint itself. In this work, we proposed a general approach to incorporate physics-based constraints into the Bayesian inversion framework, where uncertainty associated with the constraint itself can also been considered. Moreover, this idea is also extended to an approximate Bayesian approach--the ensemble Kalman filter.  

\section{Methodology}
A mathematical model of system defines a {\em forward problem} that can be formulated as
\begin{equation}
\mathbf{x} = F(\theta),
\end{equation}
where $\theta \in \mathbb{R}^{d_\theta}$ are model parameters and $\mathbf{x} \in \mathbb{R}^{d_x}$ are the states of the system. The forward operator $F$ is nominally assumed to describe a physical system, whereby $F$ typically represents a suite of algebraic and/or differential equations. In most cases, the model parameters $\theta$ are uncertain or unknown, and the state variables $\mathbf{x}$ are largely unobservable. Therefore, the unknown parameters and hidden states need to be inferred from observations $\mathbf{y}\in\mathbb{R}^{d_y}$. These observations indirectly and incompletely describe the state of the system, which can be formulated mathematically as
\begin{equation}
\mathbf{y} = H\mathbf{x} + \epsilon, 
\end{equation}
where $H$ is a projection operator projecting the full state to the observed space and $\epsilon$ represents measurement error. The standard inverse problem deals with estimating the unknown parameters $\theta$ (or the hidden states $x$) based on the observations $y$. In practice, approximate Bayesian inversion frameworks, such as Kalman filtering and Sequential Monte Carlo, are used for computational efficiency. 

%In the following subsections, we will discuss how to formulate equality and inequality constraints within the exact Bayesian framework and an approximate Bayesian filtering approach, i.e., Ensemble Kalman Filter (EnKF). 

\subsection{Constraints in Exact Bayesian Inference}
Inverse problems are typically ill-posed because the observational data is not sufficient to uniquely determine the unknown parameters. Thus, specification of additional constraints can be useful to regularize the inverse problem. Equality constraints can be defined with respect to the state variables $x$ as,
%The standard inverse problems deal with estimating the unknown parameters $\theta$ or the hidden states $x$ based on the observation $y$. However, this kind of inverse problems are usually ill-posed, i.e. the output information is not enough to uniquely determine the unknown parameters. Therefore, additional physics-based constraints need to be imposed to regularize the inverse problem. The constraint is defined with respect to the state variables $x$:
\begin{equation}
G(x) = \left[ g_1(x), g_2(x), ... , g_{d_g}(x) \right]^T = {\bf 0}\;,
\label{eq:constraint}
\end{equation}
where $g_i(x), i = 1,2, ..., d_g$ represent different equality constraints. In many application, the constraint only approximates reality. Thus, instead of directly imposing a hard constraint, we assume that each constraint satisfies a zero-mean Gaussian distribution, expressed as
\begin{equation}
G(x) \sim \mathcal{N}(\mathbf{0}, \Sigma_c) \;.
\end{equation}
where the $\Sigma_c$ is a covariance matrix used to control the strictness of each constraint. Since $x$ is intrinsically a function of $\theta$, the constraints can alternatively be expressed in terms of the parameters
\begin{equation}
G(x) = G(F(\theta)) \sim \mathcal{N}(\mathbf{0}, \Sigma_c) \;.
\end{equation}
As such, these constraints on the parameters $\theta$ can be more naturally considered within the Bayesian framework by imposing additional likelihood functions introduced by these nondeterministic constraints. 
%where $g_i(x), i = 1,2, ..., d_g$ represent different equality constraints. Because $x$ is intrinsically a function of the parameters $\theta$, the state constraints equivalently impose constraints on the parameters, i.e.

Without loss of generality, both the prior and likelihood are assumed Gaussian. Namely, the prior of the parameters $\theta$ is defined by
\begin{equation}
p(\theta) = \frac{1}{\sqrt{(2\pi)^{d_\theta} |\Sigma_\theta |}} \exp\left(-\frac{1}{2}(\theta - \hat{\theta})^T \Sigma_\theta^{-1}(\theta - \hat{\theta})\right)\;,
\label{eq:prior}
\end{equation}
where $\hat{\theta}$ and $\Sigma_\theta$ are the prior mean and covariance, which are based on existing knowledge or preliminary estimation. The observation data errors are also assumed to follow a zero-mean Gaussian distribution, i.e., $\epsilon \in N(0, \Sigma_l)$, thus the likelihood of the observed data set on $y$ is, 
\begin{equation}
p(\mathcal{D}|\theta) = \frac{1}{\sqrt{(2\pi)^{d_y} | \Sigma_l |}}\exp\left( -\frac{1}{2} (y-HF(\theta))^T \Sigma_l^{-1} (y-HF(\theta)) \right)\;.
\end{equation}
The covariance matrix $\Sigma_l$ is obtained by estimating the sample variance of the observed data sets $\mathcal{D}$. The constraints are imposed by considering the following likelihood function, 
\begin{equation}
p\left(G(x) = \mathbf{0}~|~\theta \right) = \frac{1}{\sqrt{(2\pi)^{d_g} |\Sigma_c |}} \exp\left( -\frac{1}{2} G(F(\theta))^T\Sigma_c^{-1} G(F(\theta)) \right)\;.
\end{equation}
The likelihood of the constraints defines a fitness of a specific value of $\theta$ based on the satisfaction of the constraints.
By introducing this Gaussian-type likelihood function, we enable a ``soft" enforcement of the constraints. The strictness of the constraint can be controlled by the diagonal variance matrix $\Sigma_c$,
\begin{equation}
\Sigma_c = \text{diag} \{ \sigma_{c,1}^2, \sigma_{c,2}^2,..., \sigma_{c,d_g}^2\}\;.
\end{equation}
where the variance $\sigma_i$ represent a confidence on the accuracy of the constraint. Smaller $\sigma_{c,i}$ corresponds to a stricter constraint. 

Inequality constraints can be converted to equivalent equality constraints. For example, a scalar inequality constraint $g(x) \leq { 0}$ can be expressed as 
\begin{equation}
\max\left(0, g(x) \right) = 0\;,
\end{equation}
and thus the corresponding likelihood can be expressed as 
\begin{equation}
p\left(g(x) \leq { 0}~|~\theta \right) = \frac{1}{\sqrt{2\pi \sigma_c^2 }} \exp\left( -\frac{1}{2\sigma_c^2} \left[\max\left(0, g(x) \right) \right]^2  \right)\;.
\end{equation}

Imposing constraints through a likelihood function can also be extended to disjunctive constraints. For example, consider a constraint of the form $g_1(x) = 0 \vee g_2(x) = 0$. By the union rule of probability
\begin{align}
&p\left( g_1(x) = 0 \vee g_2(x) = 0~|~\theta \right)  \nonumber \\
&= p\left( g_1(x)=0 | \theta \right) + p\left( g_2(x)=0 | \theta \right) - p\left( g_1(x) = 0 \wedge g_2(x) = 0 | \theta \right) \nonumber \\
&= \frac{1}{\sqrt{2\pi \sigma_{c,1}^2 }} \exp\left( -\frac{g_1(F(\theta))^2}{2\sigma_{c,1}^2}   \right) 
+ \frac{1}{\sqrt{2\pi \sigma_{c,2}^2 }} \exp\left( -\frac{g_2(F(\theta))^2}{2\sigma_{c,2}^2}   \right) \nonumber \\
&- \frac{1}{\sqrt{(2\pi)^{2} |\Sigma_c |}} \exp\left( -\frac{1}{2}
\begin{bmatrix}
    g_1(F(\theta)) \\
    g_2(F(\theta))
\end{bmatrix}
^T
\Sigma_c^{-1}
\begin{bmatrix}
    g_1(F(\theta)) \\
    g_2(F(\theta))
\end{bmatrix}
  \right)\;,
\end{align}
where $\Sigma_c$ again defines the covariance matrix of constraints.

With the prior distribution, likelihood of the data, and likelihood of the constraints now defined, the posterior probability distribution conditioned on the observed data $\mathcal{D}$ and the constraints $G(x)$ can be defined as 
\begin{align}
p(\theta | \mathcal{D}, G(x) = {\bf 0})   
 &= \frac{p(\mathcal{D} | \theta) p(G(x)= {\bf 0} | \theta) p(\theta)}{p(\mathcal{D}, G(x) = {\bf 0})}  \nonumber \\
& \propto p(\mathcal{D} | \theta) p(G(x)= {\bf 0} | \theta) p(\theta)\;.
\end{align}
Since the posterior distribution cannot be solved analytically in general, it is commonly evaluated based on MCMC sampling.%it is commonly approximated by MCMC sampling.

\subsection{Constraints in Approximate Bayesian Inference}
The direct Bayesian inference based on MCMC sampling is usually intractable when the likelihood calculation involves a computationally expensive model; instead approximate Bayesian approaches are commonly used to provide a more computationally tractable solution. The EnKF is one such method, which is a variant of the standard Kalman filter where the covariance matrix is replaced by Monte Carlo samples. 

For EnKF, we combine the original hidden states $x$ and the unknown parameters $\theta$ into a new augmented state 
\begin{equation}
z = \left[\theta^T , x^T \right]^T \;,
\end{equation} 
which will be updated during the filtering process according to the observed data $\mathcal{D}$. The initial ensemble is first obtained by sampling the prior distribution $p(\theta)$ and evaluating the model at each ensemble member
\begin{equation}
\left\{ \hat{z}^ {(j)}\right\}_{j=1}^J = \left\{ \left[ \hat{\theta}^{(j)}; \hat{x}^{(j)} \right]^T \right\}_{j=1}^J
= \left\{ \left[ \hat{\theta}^{(j)}; F\left( \hat{\theta}^{(j)} \right) \right]^T \right\}_{j=1}^J \;,
\end{equation} %\todo {$J$ is number of ensembles?}
%and the probability associated with each ensemble member is uniform
where $J$ is the number of ensemble members. The probability associated with each ensemble member is initially set to be uniform
\begin{equation}
w_j \triangleq p(z = \hat{z}^{(j)}) = \frac{1}{J}, ~~ j = 1,2,...,J\;.
\end{equation}
Then the expectation and covariance matrix of the state variables are estimated from the ensemble as
\begin{align}
\mathbb{E} (\hat{z}) &= \sum_{j =1}^J w_j \hat{z}^{(j)} \;, \\
\mathbb{C} (\hat{z}) &= \sum_{j =1}^J w_j (\hat{z}^{(j)}- \mathbb{E}(\hat{z}))(\hat{z}^{(j)}- \mathbb{E}(\hat{z}))^T \;.
\end{align}
If the observed data follows a normal distribution $N(\bar{y}, \Sigma_l)$, 
the prior ensemble can be updated by the observed data $\mathcal{D}$ according to the Kalman update
\begin{align}
z^{(j)} = \hat{z}^{(j)}+ \mathbb{C} (\hat{z}) H^T (H \mathbb{C} (\hat{z}) H^T + \Sigma_l)^{-1} (\bar{y} - H z^{(j)}),&  \nonumber \\
 j = 1,2,..., J\;.&
\label{eq:KF}
\end{align}
The posterior ensemble $\left\{ z^{(j)}\right\}_{j=1}^J$ represents a sampling for the posterior probability distribution $p(z | \mathcal{D})$, 
with the probability associated with each ensemble member equal to
\begin{equation}
p(z^{(j)}|\mathcal{D}) = w_j,~~\forall j = 1,2,..., J\;.
\end{equation}

Now we consider inclusion of constraints. 
The likelihood of the constraint $G(x) = {\bf 0}$ to be satisfied conditioned on each member of the posterior ensemble can be computed as
\begin{align}
Lg^{(j)} &\triangleq p\left(G(x) = {\bf 0}| z^{(j)} \right)  \nonumber \\
&= \frac{1}{\sqrt{(2\pi)^{d_g} |\Sigma_c |}} \exp\left( -\frac{1}{2} G\left(x^{(j)} \right)^T\Sigma_c^{-1} G\left(x^{(j)}\right) \right) \;.
\end{align}
By Bayes theorem, the posterior probability density of each ensemble member conditioned on the observed data $\mathcal{D}$ 
and constraints $G(x) = {\bf 0}$ is given by
\begin{align}
&p\left(z^{(j)}| \mathcal{D}, G(x) = {\bf 0} \right) = \frac{1}{Z} p\left(G(x) = {\bf 0}, z^{(j)} | \mathcal{D}\right) \nonumber \\
& = \frac{1}{Z} p\left(G(x) = {\bf 0}| z^{(j)} \right) p\left(z^{(j)}|\mathcal{D}\right) = \frac{1}{Z}w_jLg^{(j)}
\end{align}
where $Z$ is the normalization constant defined as 
\begin{equation}
Z = \sum_{j = 1}^J p\left(G(x) = {\bf 0}| x^{(j)} \right) p\left(x^{(j)}|\mathcal{D}\right) = \sum_{j = 1}^J w_j Lg^{(j)}  \;.
\end{equation}
The empirical distribution for $p\left(z | \mathcal{D}, G(x) = {\bf 0}\right) $ can be described by the posterior ensemble $\left\{ z^{(j)}\right\}_{j=1}^J$,
 and the associated probability mass %\todo{probability mass function?}
 \begin{equation}
p\left(z = z^{(j)}| \mathcal{D}, G(x) = {\bf 0} \right) = \frac{w_j Lg^{(j)}}{\sum_{p = 1}^{J} w_j Lg^{(p)}},~~\forall j = 1,2,..., J,
\end{equation}
for each ensemble member. We here re-define the new weights %\todo{why? Is it ``Thus, including constraints effectively yields a new weighting given by ...''} 
for each ensemble members as 
\begin{equation}
w_j'\triangleq \frac{w_j Lg^{(j)}}{\sum_{p = 1}^{J} w_j Lg^{(p)}}\;.
\label{eq:reweigh}
\end{equation}
The state estimation for the current iteration step is thus computed as 
 the expectation of the empirical posterior distribution conditioned on the data $\mathcal{D}$ and the prior knowledge that $G(x) = {\bf 0}$ according to
 \begin{align}
\bar{z} = \mathbb{E} \left(z | \mathcal{D}, G(x) = {\bf 0}\right) &= \sum_{j=1}^J p\left(z = z^{(j)}| \mathcal{D}, G(x) = {\bf 0} \right) z^{(j)}  \nonumber \\
&= \sum_{j=1}^J w_j' z^{(j)} \;.
\label{eq:expectation_with_constraint}
\end{align}
Then the estimation of the unknown parameters $\bar{\theta}$ can be extracted from the estimation of the full augmented state. 
%After the current step is finished, the weights of the ensemble member is updated by
%\begin{equation}
%w_j = \frac{w_j Lg^{(j)}}{\sum_{p = 1}^{J} w_j Lg^{(p)}}\;.
%\end{equation}
Also, the covariance of the parameter $\theta$ with respect to $p \left(z | \mathcal{D}, G(x) = {\bf 0}\right)$
can be computed as 
\begin{equation}
\Sigma_\theta = \sum_{i=1}^J\left( \theta^{(j)} - \bar{\theta} \right)\text{diag}\left\{ w_j' \right\}_{j =1}^J \left( \theta^{(j)} - \bar{\theta} \right)^T\;.
\end{equation}
%Before the updated weights and the posterior ensemble $\left\{ z^{(j)}\right\}_{j=1}^J$ is assigned as the new prior distribution for the next step,
%we 
The new prior ensemble for the next iteration step is obtained by sampling the following normal distribution
\begin{equation}
\hat{\theta}^ {(j)} \sim N(\bar{\theta}, \Sigma_\theta), ~~j = 1,2,..., J\;,
\end{equation}
to maximize the next step prior entropy \cite{penfield2010principle} while keeping the mean and covariance the same as the previous posterior distribution.
The iterative process continues until a stopping criterion is satisfied or the maximum iteration number is reached.

%The updated weights and the posterior ensemble $\left\{ z^{(j)}\right\}_{j=1}^J$ is assigned as the new prior distribution to continue the next iteration step. 
%The convergence of hidden states estimation is checked at each iteration step as the stopping criterion.

%\section{[TODO applying Gaussian kernel to provide robustness against outlier]}

\section{Results and Discussion}

\subsection{Model Test Problem}
To verify the effectiveness of the constrained Bayesian inference framework described above, a simple test case is presented here. 
The forward model mapping from the parameter space $\Theta \subset  \mathbb{R}^2$ to the state space $X \subset \mathbb{R}^2$ is defined as 
\begin{equation}
\begin{bmatrix}
    x_1   \\
    x_2
\end{bmatrix}
=
F(\theta)
=
\begin{bmatrix}
     \exp(-(\theta_1 + 1)^2 - (\theta_2 + 1)^2)   \\
     \exp(-(\theta_1 - 1)^2 - (\theta_2 - 1)^2)
\end{bmatrix}\;.
\end{equation}
%\begin{align}
%x_1 &= \exp(-(\theta_1 + 1)^2 - (\theta_2 + 1)^2) \nonumber \\
%x_2 &= \exp(-(\theta_1 - 1)^2 - (\theta_2 - 1)^2)
%\end{align}
The projection matrix mapping from state space to output is given by
\begin{equation}
H = [-1.5, -1.0]\;,
\end{equation} %\todo{no noise?}
and thus the reconstructed output is 
\begin{align}
HF(\theta)&= -1.5 \exp(-(\theta_1 + 1)^2 - (\theta_2 + 1)^2)  \nonumber \\
&~~~   - 1.0 \exp(-(\theta_1 - 1)^2 - (\theta_2 - 1)^2)\;, 
\end{align}
where $HF(\theta) \in \mathbb{R}^1 $. We consider the following constraint:
\begin{equation}
G(x) = - 0.25 \log x_1 + 0.25 \log x_2 -2 = 0\;,
\end{equation}
which can be equivalently written in terms of $\theta$,
\begin{equation}
G(F(\theta)) = \theta_1 + \theta_2 - 2 = 0\;.
\label{eq:constraint_theta}
\end{equation}
We assume the observed data follow the normal distribution $N(\bar{y}, \Sigma_l)$ where the mean $\bar{y} = -1.0$ 
and the covariance matrix $\Sigma_l $ is chosen based on the uncertainty associated with data.

%[TODO] Plot the cost function with respect to the parameter space, explain the reason why we want to set up the problem this way is to introduce multiple local minimum, 
%so that I have to use the constraint to help to converge to the true minimum.

\begin{figure}
    \centering
    \includegraphics[width=0.45\columnwidth]{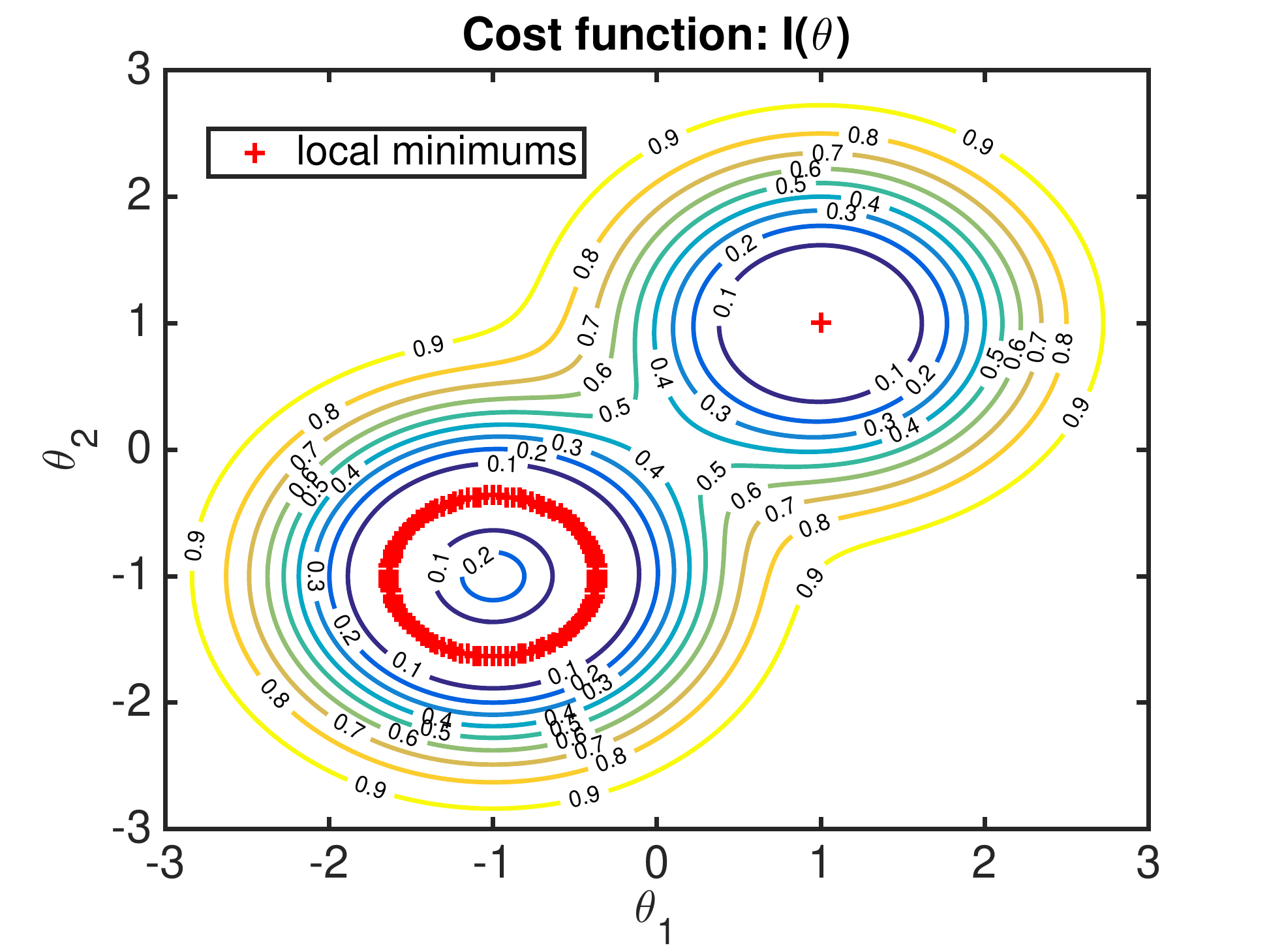}
    \caption{Contour plot of the cost function $I(\theta)$ with respect to parameters $\theta = \left[ \theta_1, \theta_2 \right]$
    The red ``+" denote the local minima of the cost function.}
    \label{fig:cost_function}
\end{figure}

This model is chosen to create a simple scenario with multiple local minimums.
Namely, regardless of the prior information and constraints, 
we seek the model parameters that minimize the difference between the observed output and reconstructed output, 
quantified by the cost function 
\begin{align}
I(\theta) &= \left\|\bar{y} - HF(\theta) \right\|^2  \nonumber \\
&=  \left(1.5 \exp(-(\theta_1 + 1)^2 - (\theta_2 + 1)^2)\right.  \nonumber \\
&~~~ \left.+ 1.0 \exp(-(\theta_1 - 1)^2 - (\theta_2 - 1)^2) -1.0 \right)^2\;.
\end{align}
which has minimums at (a) $\theta^* = (1,1)$ ; and (b) $\theta$ on the circle defined by 
\begin{equation}
(\theta_1 + 1)^2 + (\theta_2 + 1)^2 = \log 1.5\;.
\end{equation}
The contour plot of the cost function and the local minimums are visualized in Figure \ref{fig:cost_function}.  
Here we assume $\theta^* = (1,1)$ is the true value of the parameter $\theta$, 
and the constraint \eqref{eq:constraint_theta} will help to eliminate convergence to other local minima.
%------------------------------------------------------------
\begin{figure}
  \centering
  \begin{tabular}{cc}
    \includegraphics[width=0.45\columnwidth]{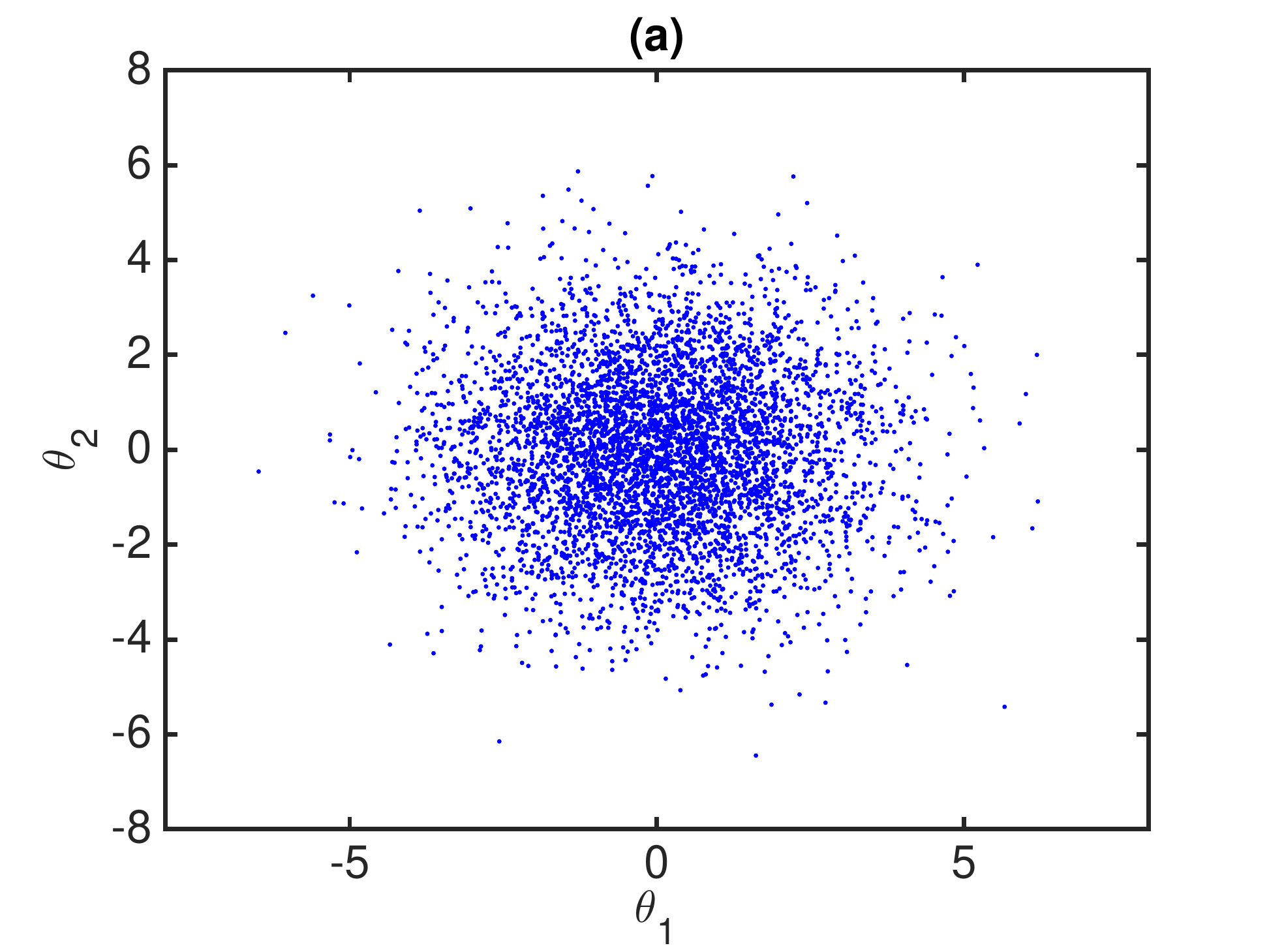}&
    
    \includegraphics[width=0.45\columnwidth]{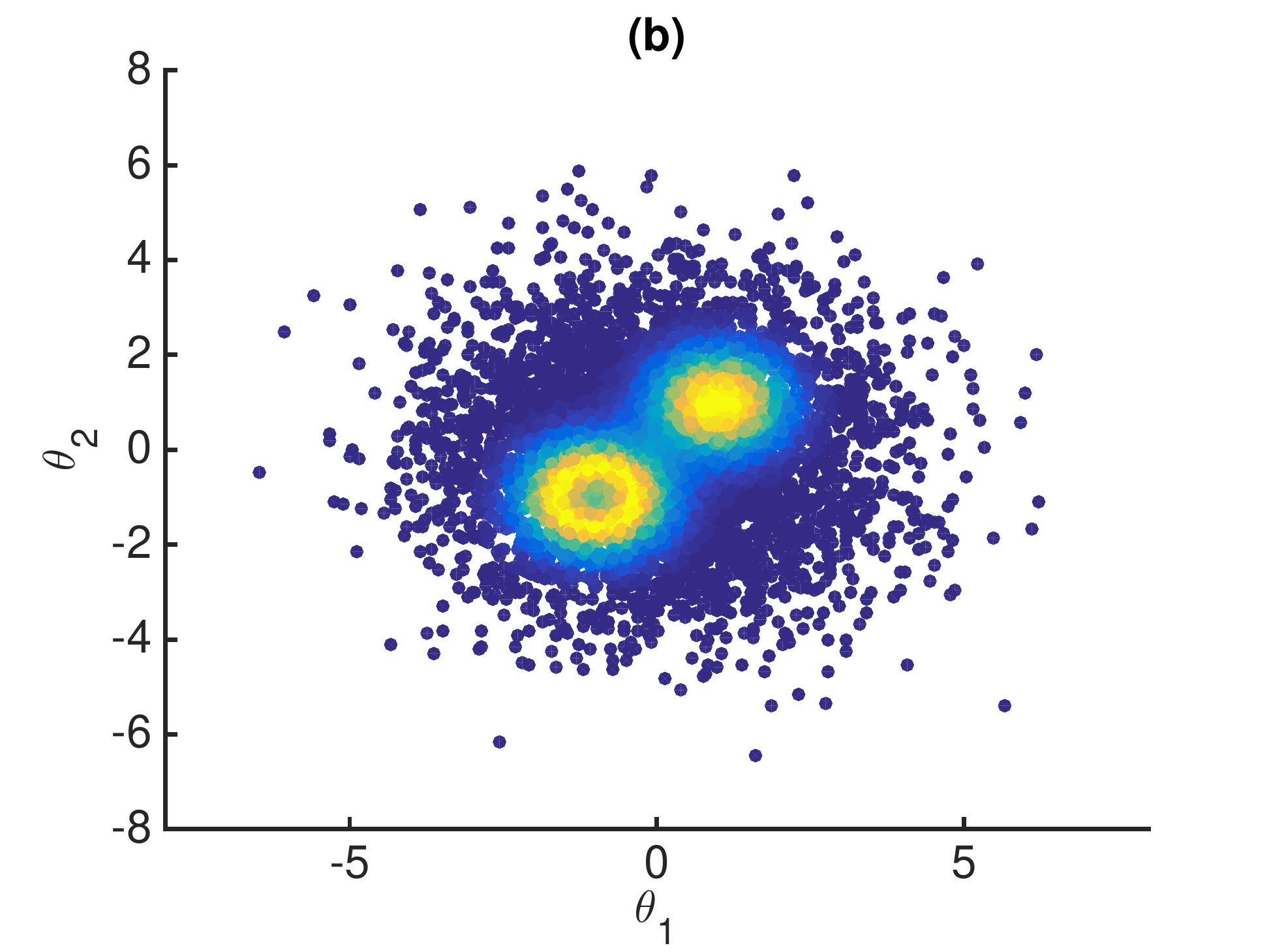}\\
    
    \includegraphics[width=0.45\columnwidth]{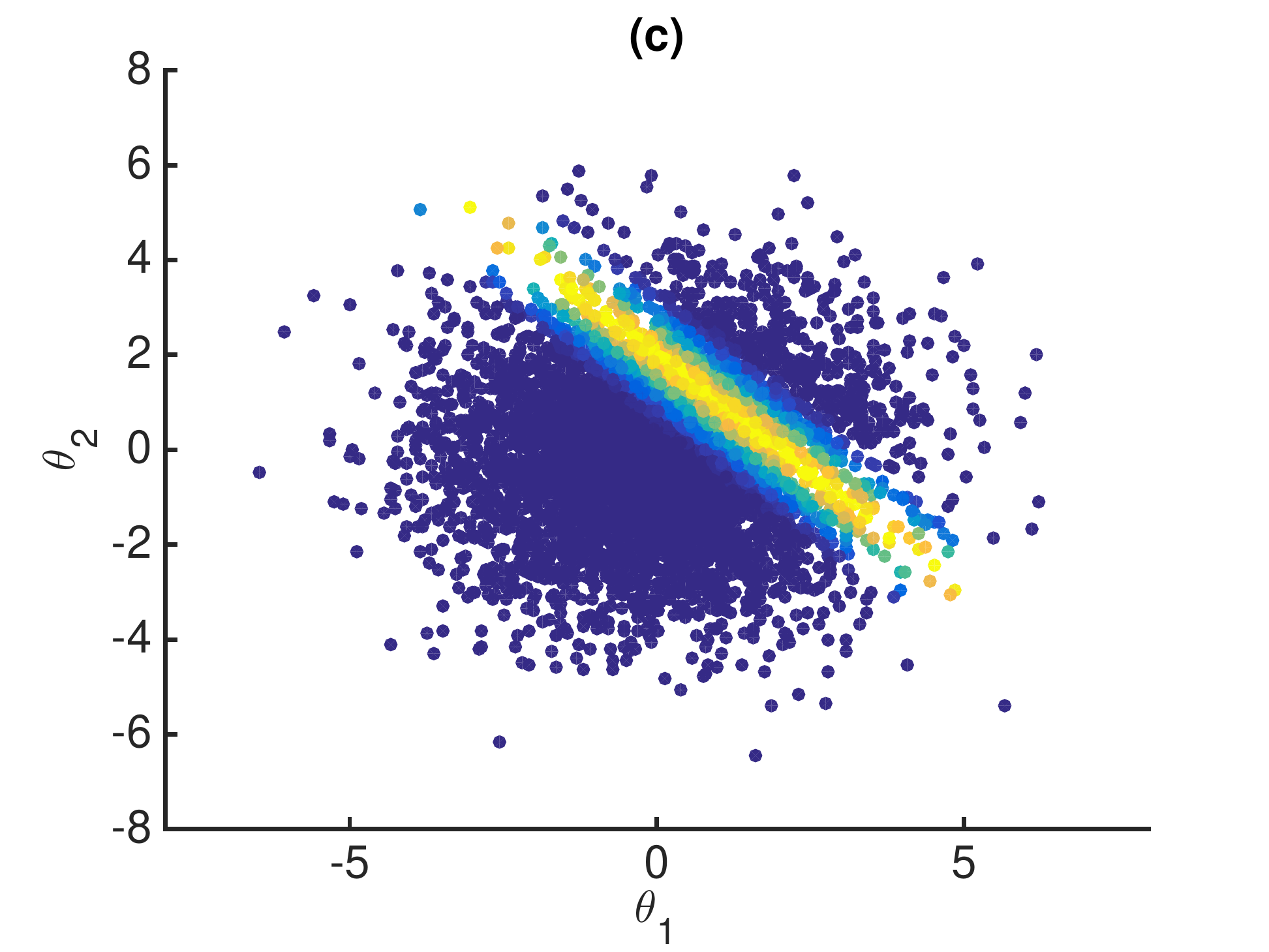}&
    
    \includegraphics[width=0.45\columnwidth]{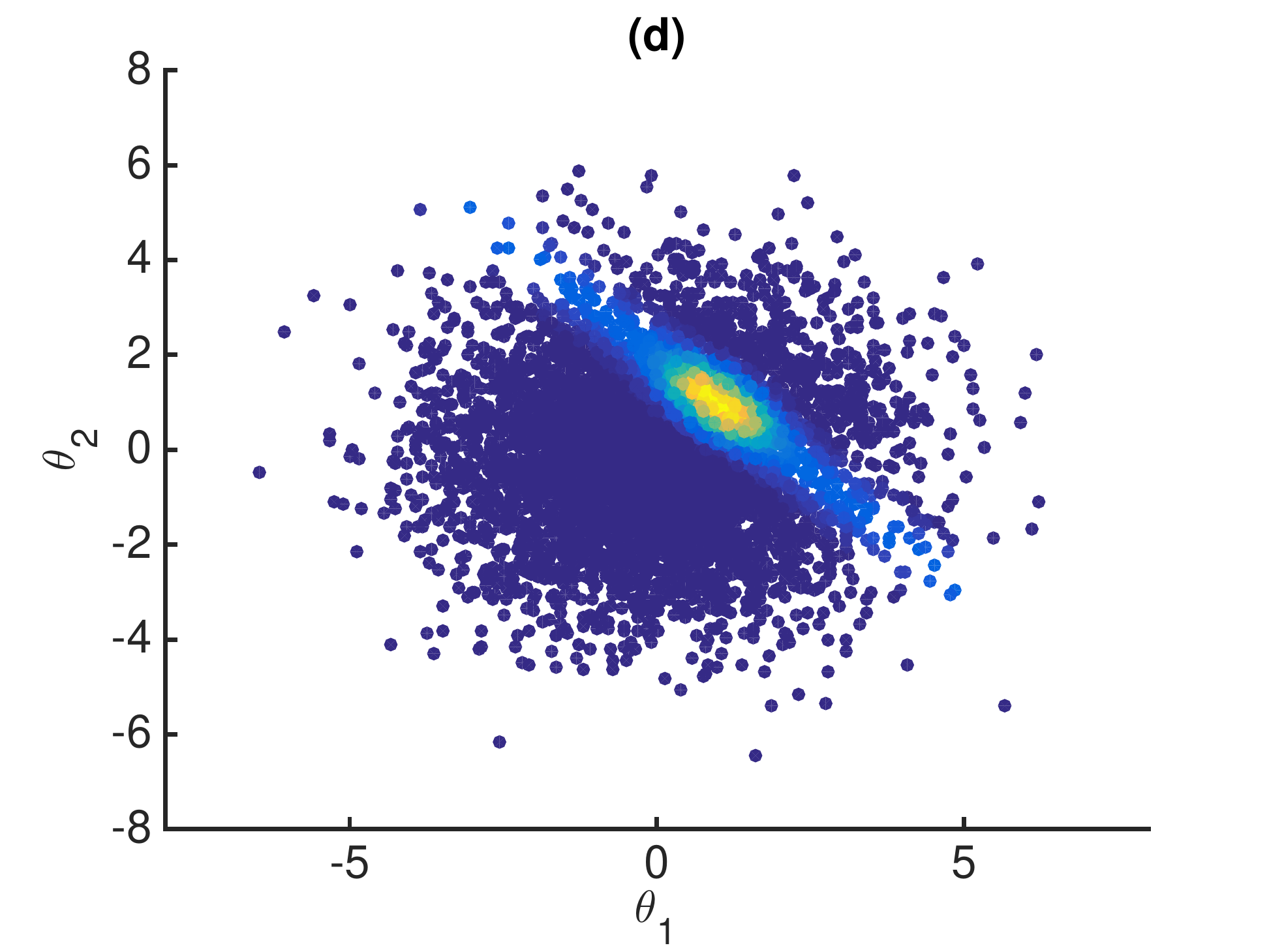}\\
  \end{tabular}
  \caption{(a) Sampling of the prior distribution; 
  		(b) Sample + data likelihood $p(\mathcal{D} | \theta)$;
		(c) Sample + constraint likelihood $p(G(x) = {\bf 0} | \theta)$;
		(d) Sample + posterior distribution $p(\theta | \mathcal{D}, G(x) = {\bf 0})$}
\label{fig:fully_bayesian}
\end{figure}
%------------------------------------------------------------

The situation of multiple local minima is a common challenge in solving inverse problems, 
because the observed output information is usually not enough to uniquely determine the unknown parameters.
We show below that imposing constraints can help the solution to converge to the true value. 

\subsection{Exact Bayesian Inference}

%------------------------------------------------------------
\begin{table} %\label{tb:fully_bayesian}
\centering
\scriptsize
\begin{tabularx}{\columnwidth}{cccccc}
\toprule
                    & $\theta_1$ & $\theta_2$  & $x_1$  &  $x_2$ & $y$ \\
\midrule
True values                        & 1            & 1           & 0          & 1            & -1 \\
\midrule
No constraint                     & -0.0448  & -0.0722 & 0.1698 & 0.1063   & -0.3610 \\
\midrule
With constraints (EXP) & 0.9926   & 0.9449   & 0.0004  &  0.9969 & -0.9976 \\
\midrule
With constraints (MAP) & 0.9845   & 0.9698   & 0.0004  &  0.9988 & -0.9994 \\
\bottomrule
%\label{tb:fully_bayesian}
\end{tabularx}
\caption{Parameters $\theta$, states $x$ and output $y$ estimated using sample-based Bayesian inference with no constraint imposed and with constraint imposed. } %\todoJW{$\tau_{\theta}$}}
\label{tb:fully_bayesian}
%\label{tab:parameters}
\end{table}
%------------------------------------------------------------

The prior distribution \eqref{eq:prior} is first sampled with $J = 5000$ samples. 
The mean and the covariance matrix are set to 
\begin{equation}
\hat{\theta} = 
\begin{bmatrix}
   0 \\
   0  
\end{bmatrix}
, ~~
\Sigma_\theta
=
\begin{bmatrix}
   3       & 0 \\
   0       & 3
\end{bmatrix}
\;.
\end{equation}
The distribution of the samples is visualized in Figure \ref{fig:fully_bayesian}(a).
The trivial zero mean  represents non-informative prior knowledge of $\theta$,
and the large variance defined by $\Sigma_\theta$ above denotes large uncertainty about the prior.
Ideally if better prior knowledge exists, we can specify a better prior here, with more accurate mean and less uncertainty.
After sampling the prior, the likelihood function of data $p(\mathcal{D}| \theta)$ is evaluated at each individual sample point $\left\{ \theta^{(j)} \right\}_{j=1}^J$.
The likelihood of the data is plotted with respect to each sample in Figure \ref{fig:fully_bayesian}(b).
The value of the likelihood is indicated by the brightness of the sample. 
It can be clearly observed that the brightest regions coincide with the local minimums in Figure \ref{fig:cost_function},
which shows the region of the highest likelihood of data .
Similarly, we evaluate the likelihood function of the constraint $p(G(x) = {\bf 0} | \theta)$ at each sample, 
and the likelihood is visualized in Figure \ref{fig:fully_bayesian}(c). 
The region with the highest likelihood represents the form of the constraint in $(\theta_1, \theta_2)$ space, 
which is $\theta_1 + \theta_2 - 2 = 0 $.
The variance for the constraint is set to be $\Sigma_c = 0.5$, which controls how strict the constraint is enforced.
Lastly, the posterior $p(\theta | \mathcal{D}, G(x) = {\bf 0})$ is evaluated at each samples, 
and the distribution of the sample along with the posterior weights are plotted in Figure \ref{fig:fully_bayesian}(d). 
It is clearly observed that the location with the highest posterior density correspond to 
the intersection between the regions with high likelihood of data $\mathcal{D}$ and high likelihood of the constraint satisfaction.
This intersection region picks out the true value of the parameter $\theta$.
Computing the weighted sum of the parameter samples  $\left\{ \theta^{(j)} \right\}_{j=1}^J$ 
with respect to the posterior weights yields the final estimation of the unknown parameter %$\theta^*_{\mbox{Exp}}$.
%\begin{align}
%\theta^*_{\mbox{MAP}} & = \arg\max_{\theta^{(j)}} p\left( \theta^{(j)} | \mathcal{D}, G(x) = {\bf 0} \right) =  \arg\max_{\theta^{(j)}} w_j \;,\\
$\theta^*_{\mbox{Exp}}  = \sum_{j=1}^J p\left( \theta^{(j)} | \mathcal{D}, G(x) = {\bf 0} \right)  \theta^{(j)} =\sum_{j=1}^J w_j \theta^{(j)}\;.$
%\end{align}
Or simply taking the sample $\theta^{(j)}$ that maximize the posterior $p\left(\theta^{(j)}| \mathcal{D}, G(x) = {\bf 0} \right) $
yields the maximum a posteriori estimation (MAP) of the unknown parameter, i.e. $\theta^*_{\mbox{MAP}}$. 
Once the parameter is estimated, the estimated value of the state variables $x$ and output $y$ can be computed by evaluating the forward model $F(\theta^*)$.
These estimated values are listed in Table \ref{tb:fully_bayesian} for the case of including and not including constraint. 
It can be seen from this table that imposing the constraint significantly increase the estimation accuracy in the case where multiple local minimums exist.

\begin{figure} [h]
    \centering
    \includegraphics[width=0.45\columnwidth]{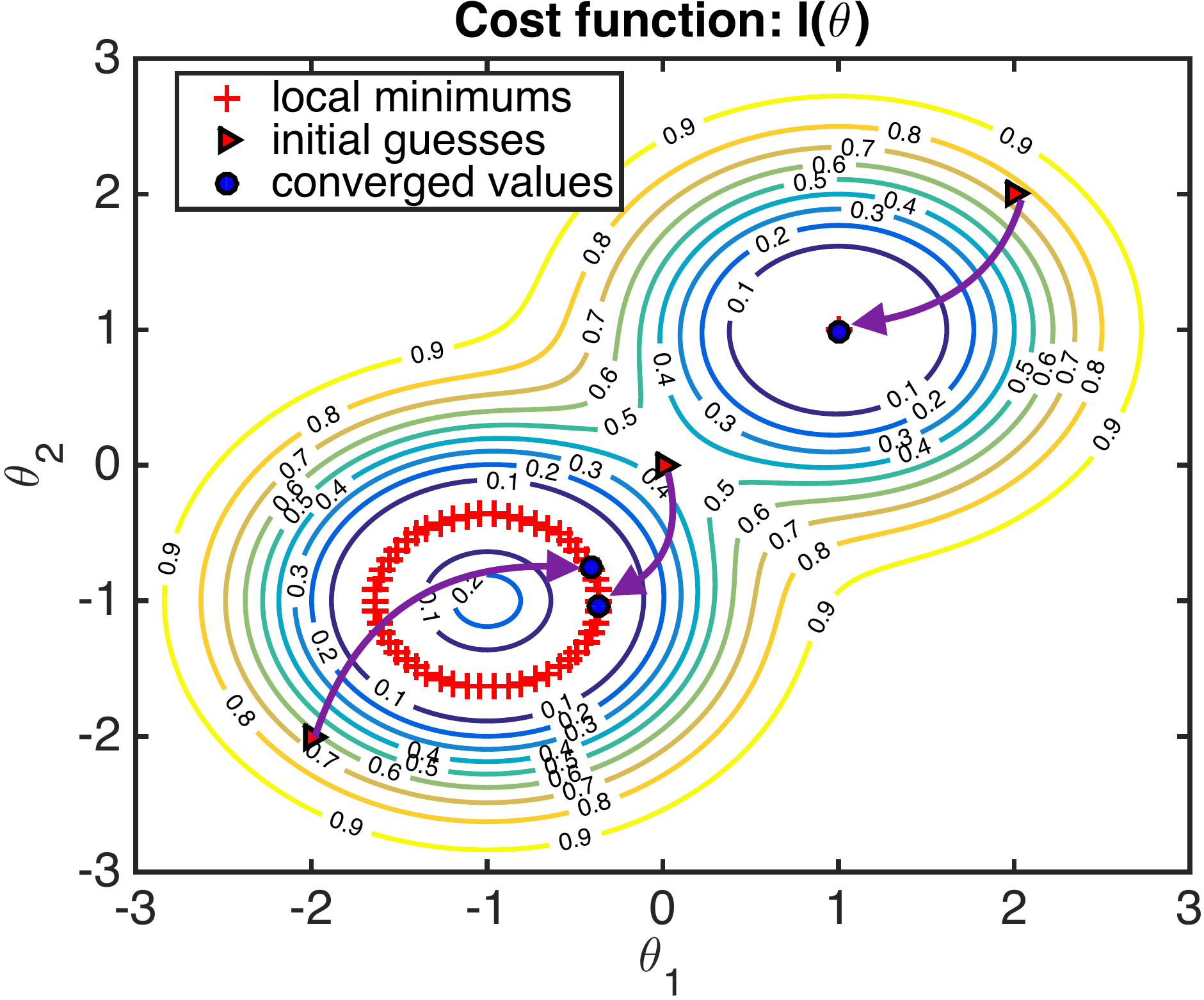}
    \caption{Different initial guesses of the unknown parameter $\theta$ (marked with red triangles) and their corresponding converged values after 1000 EnKF iteration steps (marked with blue circles), with no constraint imposed.}
    \label{fig:diff_init_no_constraint_sigma_1}
\end{figure}

\subsection{Approximate Bayesian Inference}
\subsubsection{No constraint imposed.}
Iterative ensemble Kalman filter estimates the unknown model parameters $\theta$ in a iterative manner. 
Since the cost function $I(\theta)$ has multiple local minimums, different initial guesses of $\theta$ will converge to different local minimums.
We here define 
\begin{align}
\text{Group I}  &\triangleq \left\{ \theta^* \in \mathbb{R}^2 | \theta^*  = (1,1)  \right\} \;, \\
\text{Group II} &\triangleq \left\{ \theta^* \in \mathbb{R}^2 | (\theta_1^* + 1)^2 + (\theta_2^* + 1)^2 = \log 1.5  \right\}\;,
\end{align}
which represent two different local minimum regions. 
$\theta^*$ represents the converged value of the parameter after ensemble Kalman filter iterations.

%evolution of ensemble------------------------------------------------------------
\begin{figure*}[h!]%[htp]
  \centering
  \begin{tabular}{ccc}
    \includegraphics[width=0.3\textwidth]{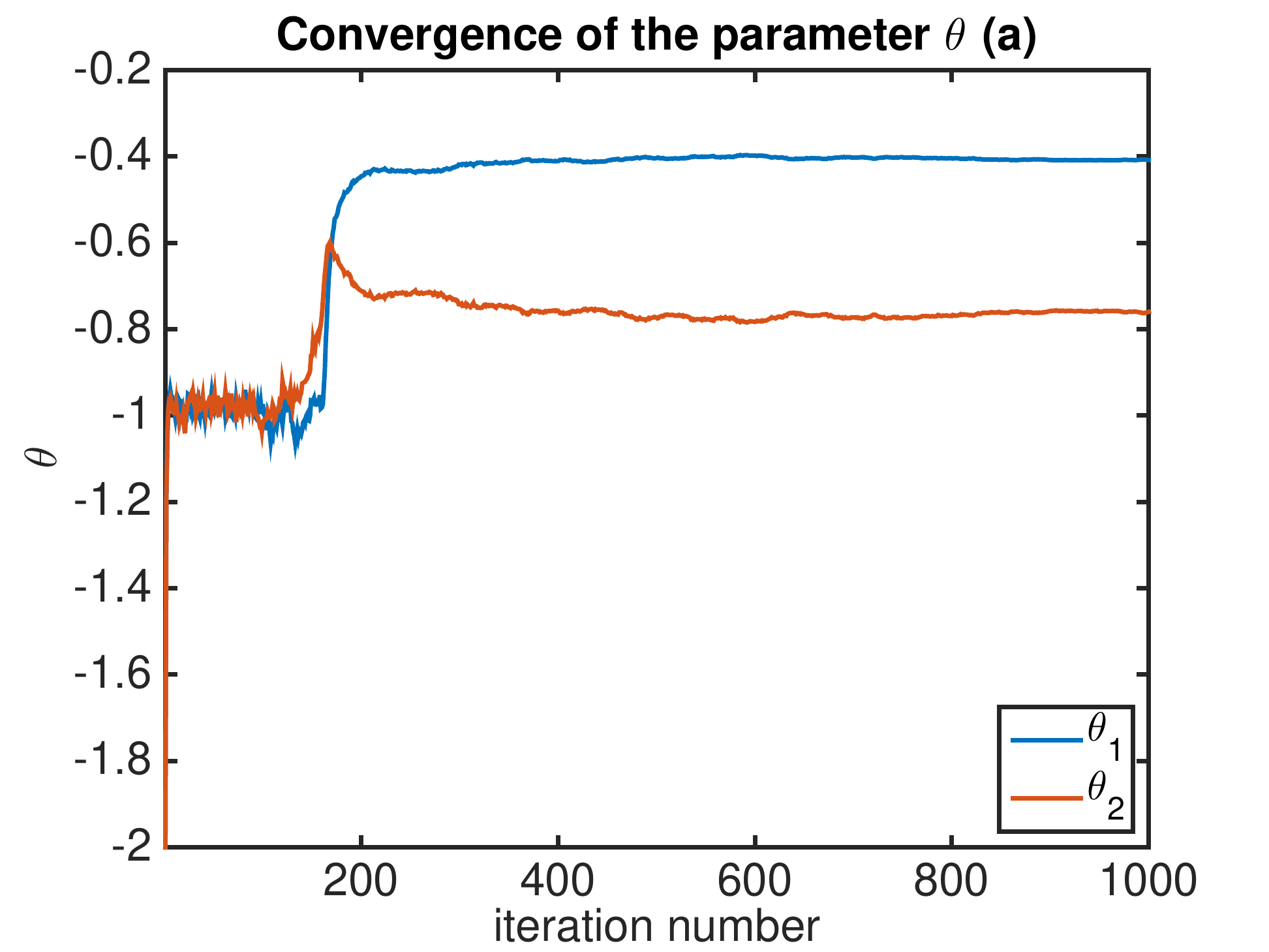}&
    
    \includegraphics[width=0.3\textwidth]{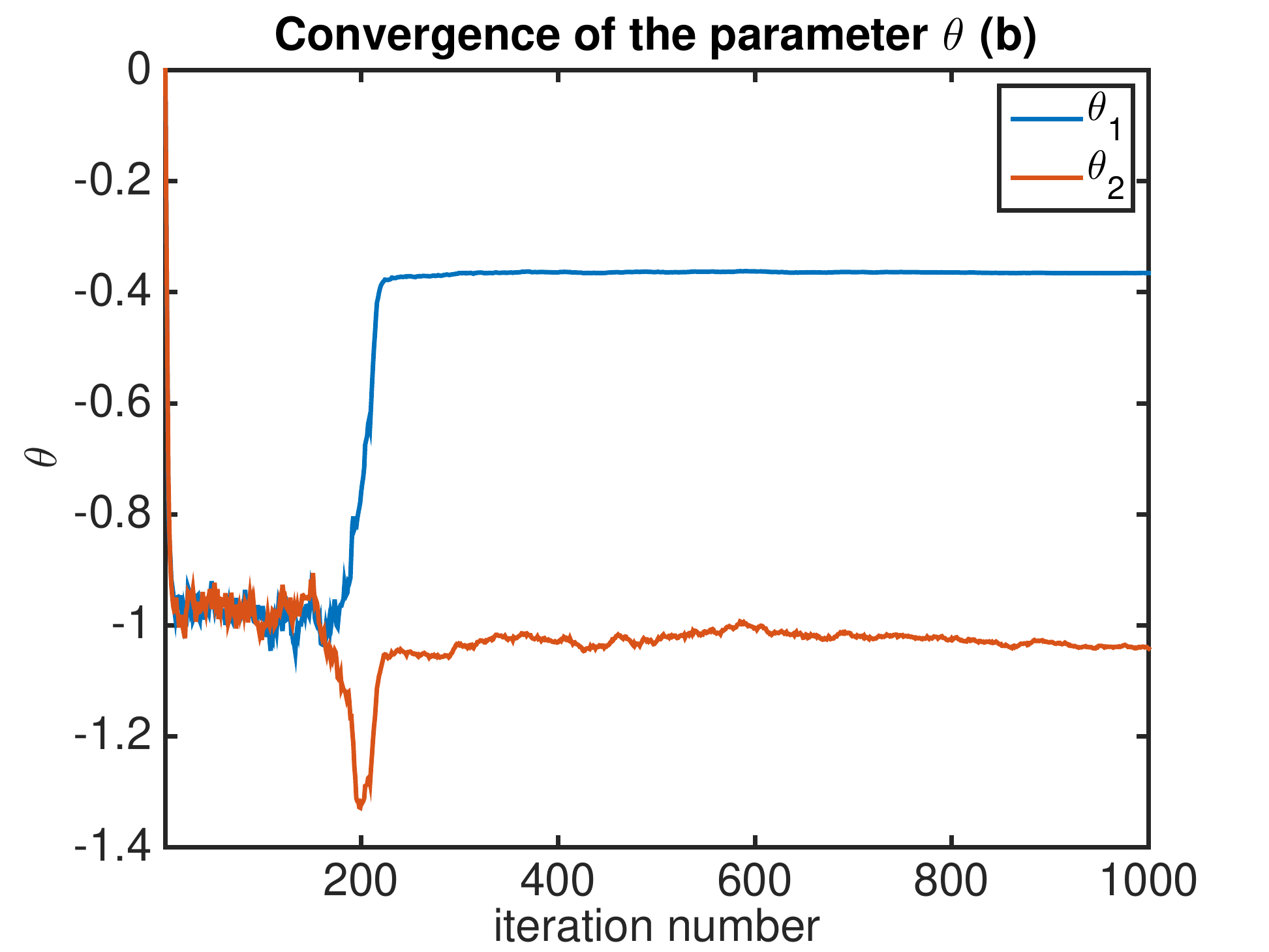}&
    
    \includegraphics[width=0.3\textwidth]{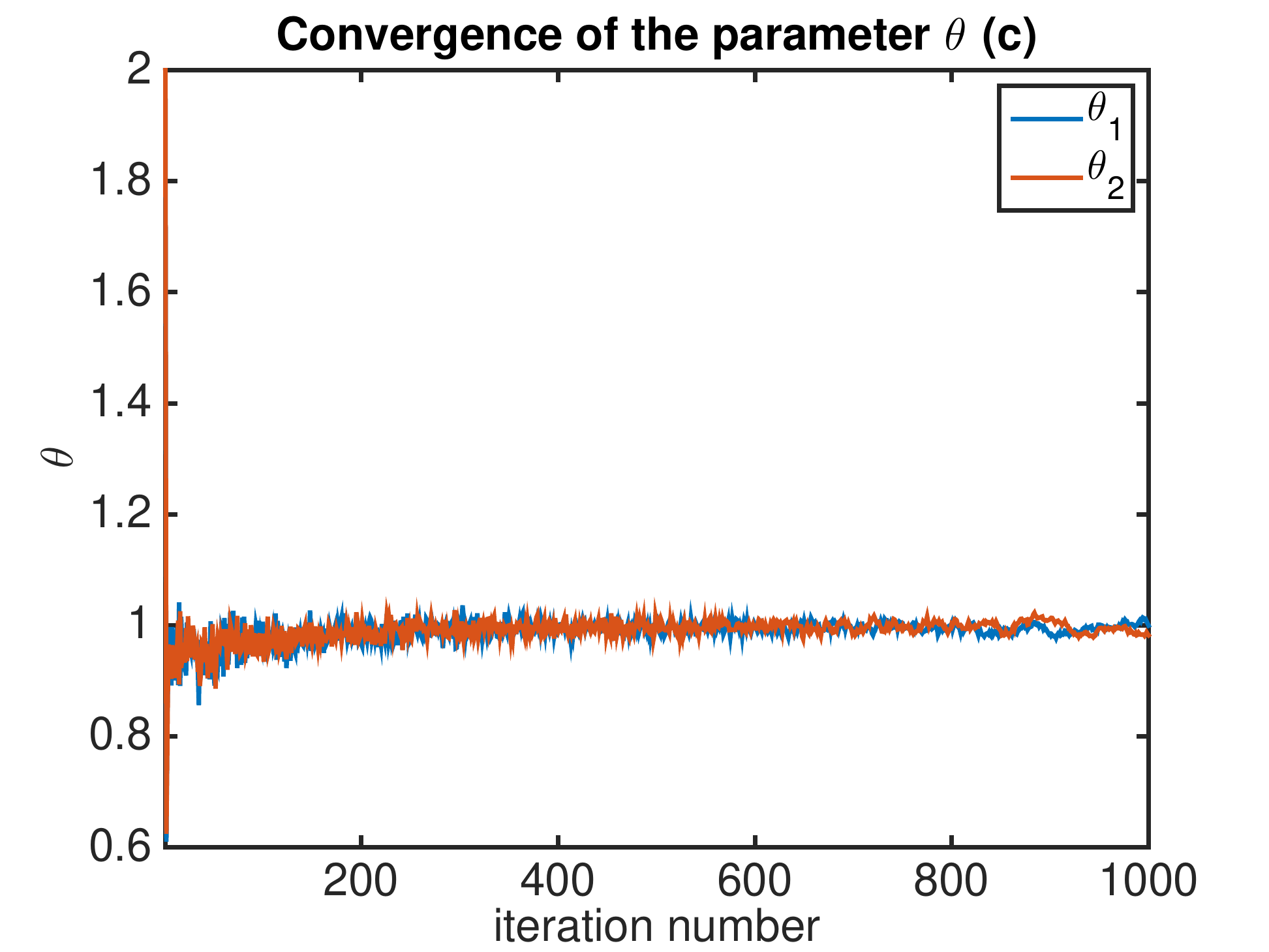}\\
  \end{tabular}
  \caption{Results for the convergence of the parameter $\theta$ with no constraint imposed.
  		(a) Initial guess $\theta^0 = (-2, -2)$; 
  		(b) initial guess $\theta^0 = (0,0)$;
		(c) initial guess $\theta^0 = (2,2)$.}
\label{fig:theta_convergence_no_constraint_sigma_1}
\end{figure*}

%evolution of ensemble------------------------------------------------------------
\begin{figure*}[h]%[htp]
  \centering
  \begin{tabular}{ccc}
    \includegraphics[width=0.3\textwidth]{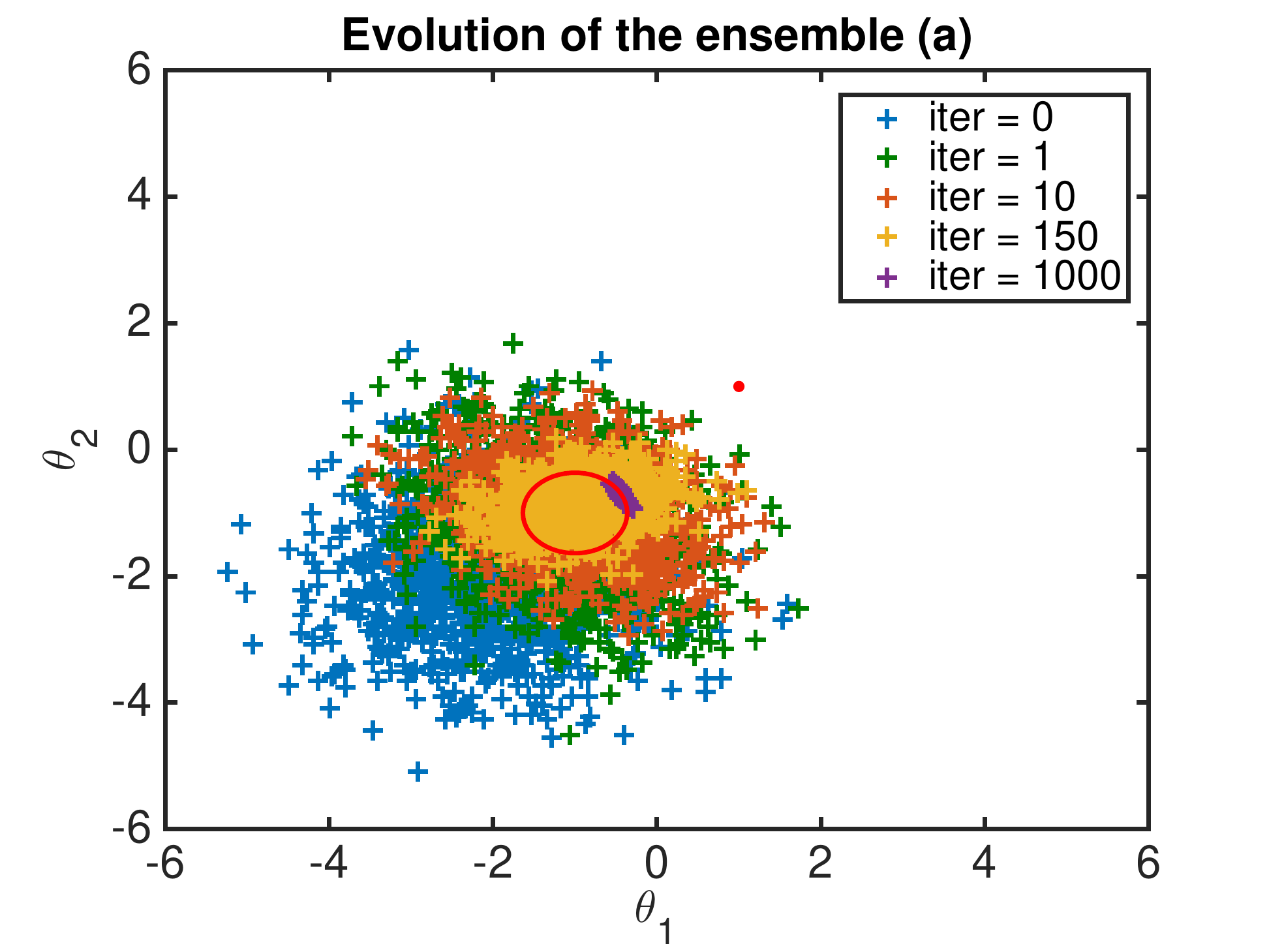}&
    
    \includegraphics[width=0.3\textwidth]{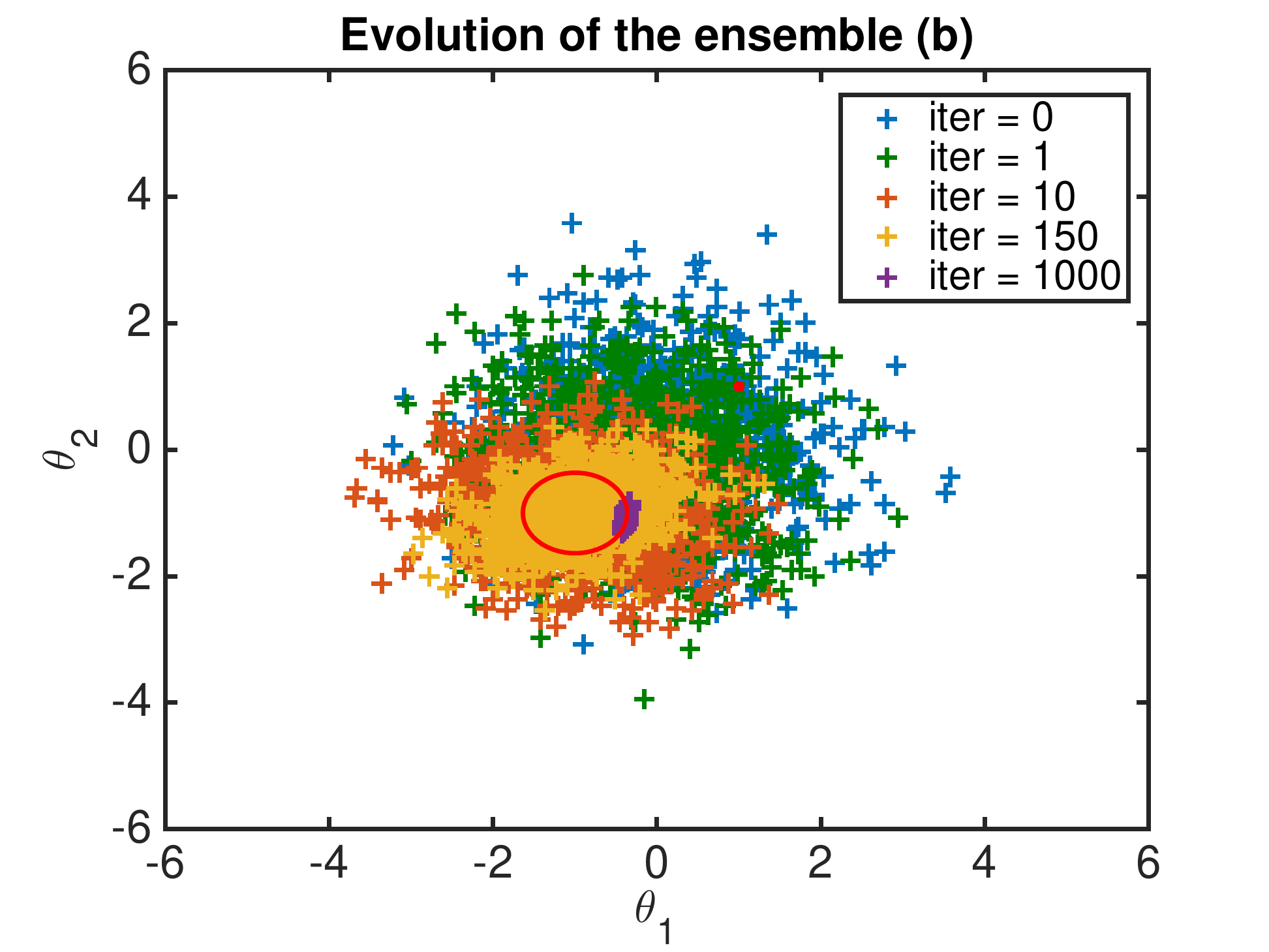}&
    
    \includegraphics[width=0.3\textwidth]{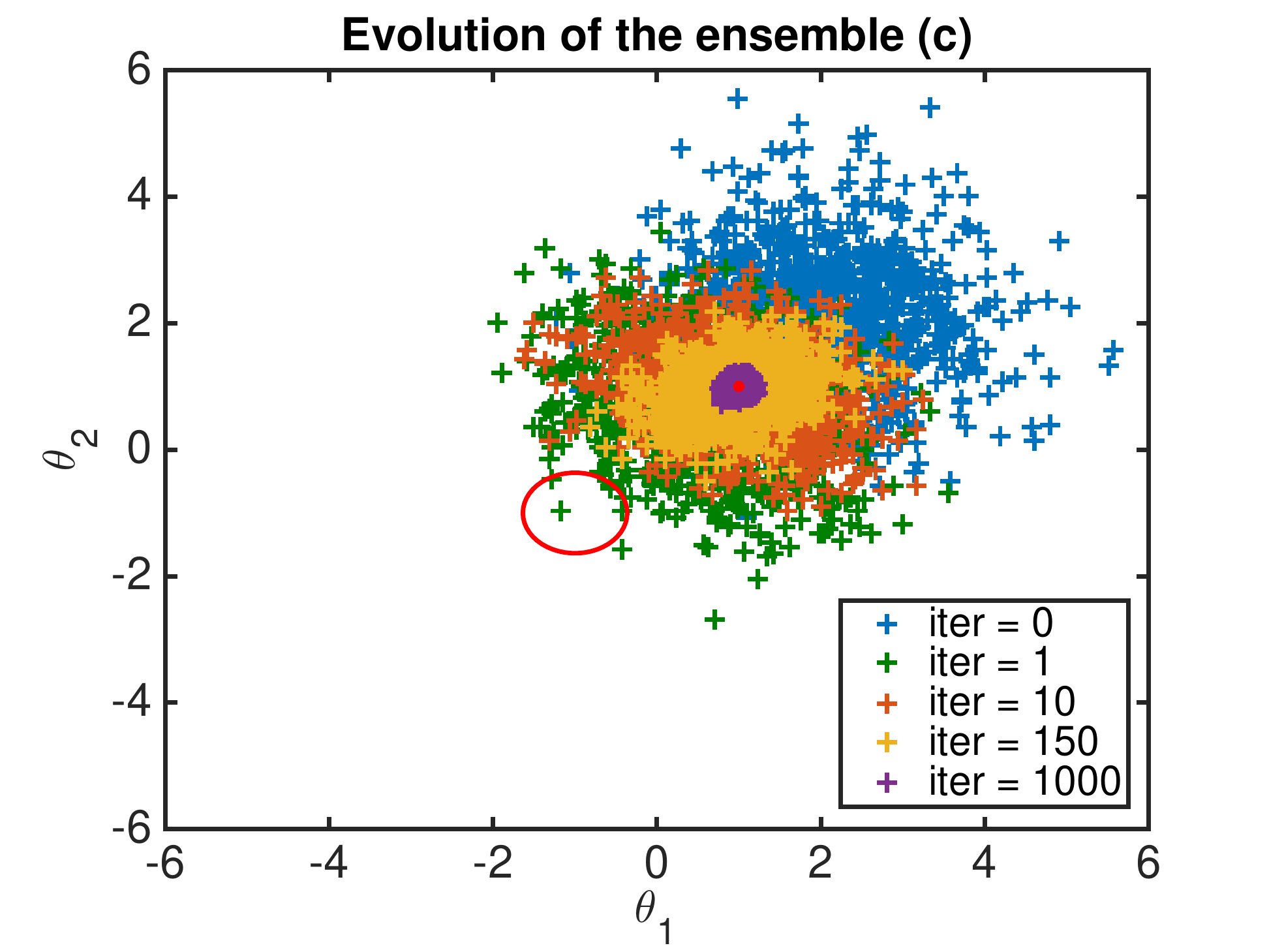}\\
  \end{tabular}
  \caption{Evolution of the ensemble of the parameter $\theta$ with no constraint imposed. 
  		The points on the red circle centered at (-1,-1) and the red point at (1,1) denote the local minimums of the cost function $I(\theta)$.
  		(a) Initial guess $\theta^0 = (-2, -2)$; 
  		(b) initial guess $\theta^0 = (0,0)$;
		(c) initial guess $\theta^0 = (2,2)$.}
\label{fig:ensemble_no_constraint_sigma_1}
\end{figure*}

Here we simulated three different cases with different initial guesses: 
		(a)  $\theta^0 = (-2, -2)$; 
  		(b)  $\theta^0 = (0,0)$;
		(c)  $\theta^0 = (2,2)$.
The covariance matrix of the prior and the covariance of the data likelihood are given as $\Sigma_\theta = [1, 0; 0, 1]$ and $\Sigma_l = 0.01$.
The results for the three different simulations are visualized in the parameter space of $\theta$ in Figure \ref{fig:diff_init_no_constraint_sigma_1}.
It can be seen that the upper right initial guess at $(2,2)$ converges to local minimum Group I,
and the other two initial guesses both converge to local minimum Group II.
The convergence processes of the parameter $\theta$ for the three different initial guesses are plotted in Figure \ref{fig:theta_convergence_no_constraint_sigma_1}.
It can be seen that all converge to the corresponding local minimum groups within about 400 iterations. 
The main difference is that while Case (c) converges to the local minimum $(1,1)$, i.e., Group I, directly after a few iterations, Case (a) and (b) converge to $\theta = (-1, -1)$ first, which is the center of the local minimum circle of Group II, and then shift to a local minimum on the circle of Group II at around the 200th iteration (indicated by the ``jump").
The reason behind this is that we use the mean of the ensemble of each step as the estimated parameter value.
When the ensemble converges to the local neighborhood of Group II, the mean of the ensemble will generally be the center of the local minimum circle because the high likely ensemble members are roughly symmetrically distributed around the center $(-1, -1)$. 
The mean of the ensemble will gradually shift to certain points on the circle based on the distribution of the ensemble members. 
The evolution of the ensemble for different initial guesses is shown in Figure \ref{fig:ensemble_no_constraint_sigma_1}. 
It can be seen that the variance of the ensemble gradually decrease until all ensemble members collapse to the corresponding local minimum.

The convergence results for the three different initial guesses are summarized here,
\begin{align}
\theta^0 = (-2, -2 ) &\rightarrow \theta^* = (-0.4080, -0.7598) \in \text{\small Group II}\;, \nonumber \\
\theta^0 = (0, 0) &\rightarrow \theta^* = (-0.3654, -1.0421) \in \text{\small Group II} \;, \nonumber \\
\theta^0 = (2, 2 ) &\rightarrow \theta^* = (0.9998, 0.9812) \in \text{\small Group I} \;. \nonumber
\end{align}
The reconstructed outputs $HF(\theta^*)$ for the above cases all converge to the target value $\bar{y} = -1$ within 1000 EnKf iterations.
However, with no constraint is imposed, the estimate of the parameter $\theta$ will converge to the closer local minimum group based on where the initial guesses are.
The initial guess in the middle $(0,0)$ converges to the local minimum Group II because Group II contains more local minimums than Group I, 
and therefore the the solution is more likely to converge to Group II when initial guess is in the middle.
More broadly, there is no guarantee that the estimate of the parameter will converge to the the true parameter value $(1,1)$.

%------------------------------------------------------------
%\begin{table}%[htbp]
%\centering
%\scriptsize
%\begin{tabularx}{\columnwidth}{cccc}
%\toprule
%$\theta^0$                   & (-2, -2)                           & (0, 0)                       & (2, 2)\\
%\midrule
%$\theta^*$                    &  (-0.4080, -0.7598)         & (-0.3654, -1.0421)    & (0.9998, 0.9812) \\
%\midrule
%$HF(\theta^*)$             & -1.0032                      &-1.0028                        &  -0.9869     \\
%\midrule
%Group   & II                                  &  II                               & I       \\
%\bottomrule
%%\label{tb:fully_bayesian}
%\end{tabularx}
%\caption{Simulation results with no constraint imposed for different initial guesses with the prior covariance $\Sigma_\theta = [1, 0; 0, 1]$, 
%    and the covariance of the data likelihood $\Sigma_l = 0.01$.} %\todoJW{$\tau_{\theta}$}}
%\label{tab:different_initial_no_constraint_sigma1}
%\end{table}
%--------------------------------------------------------

\subsubsection{With constraint imposed}
For the local minimums in Group I and Group II, only the true parameter value $(1,1)$ satisfies the constraint. 
We test here whether imposing the constraint can help the convergence of the parameter estimation to the true value.

\begin{figure*}[h]%[htp]
  \centering
  \begin{tabular}{ccc}
    
    \includegraphics[width=0.3\textwidth]{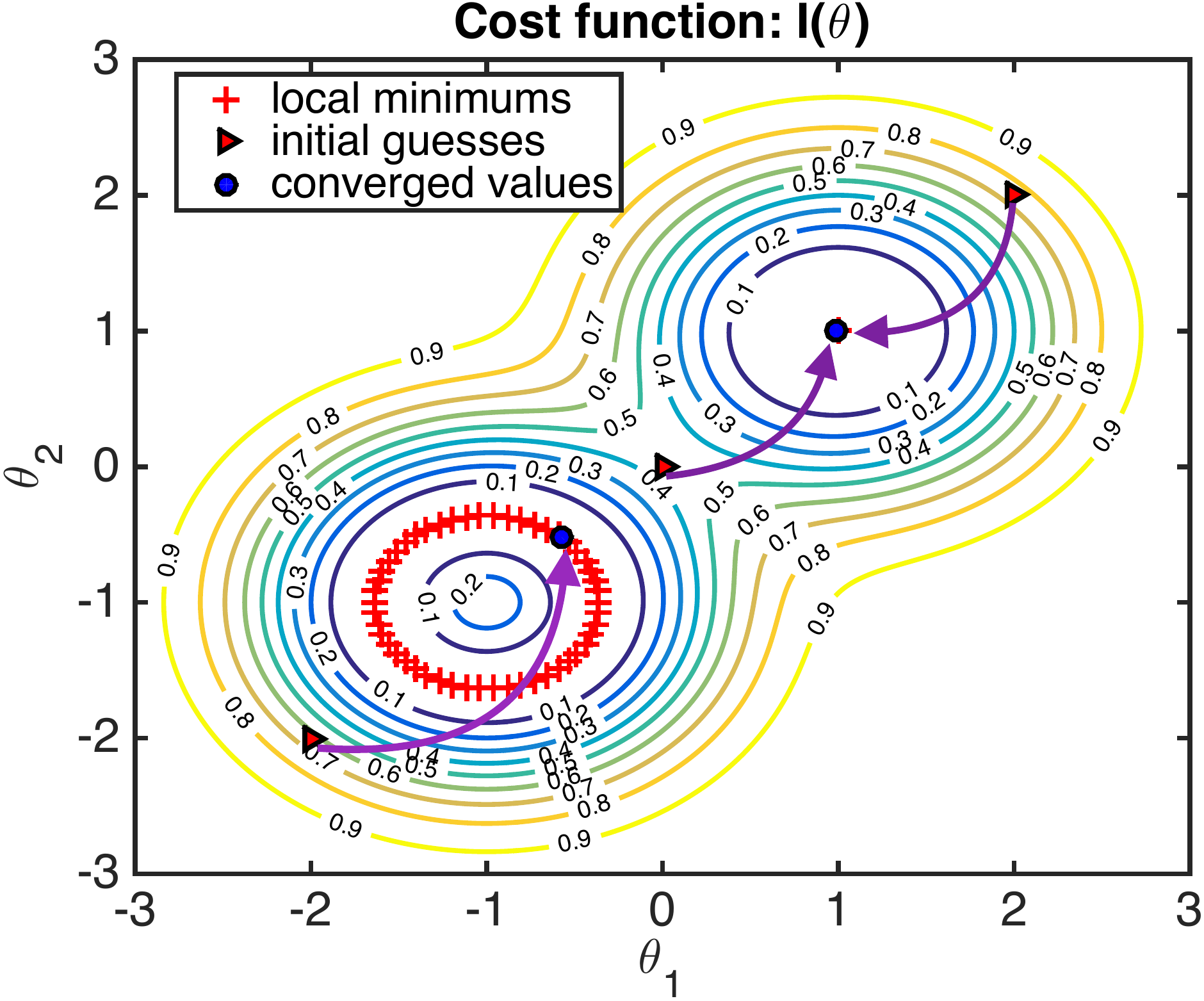}&
    
    \includegraphics[width=0.3\textwidth]{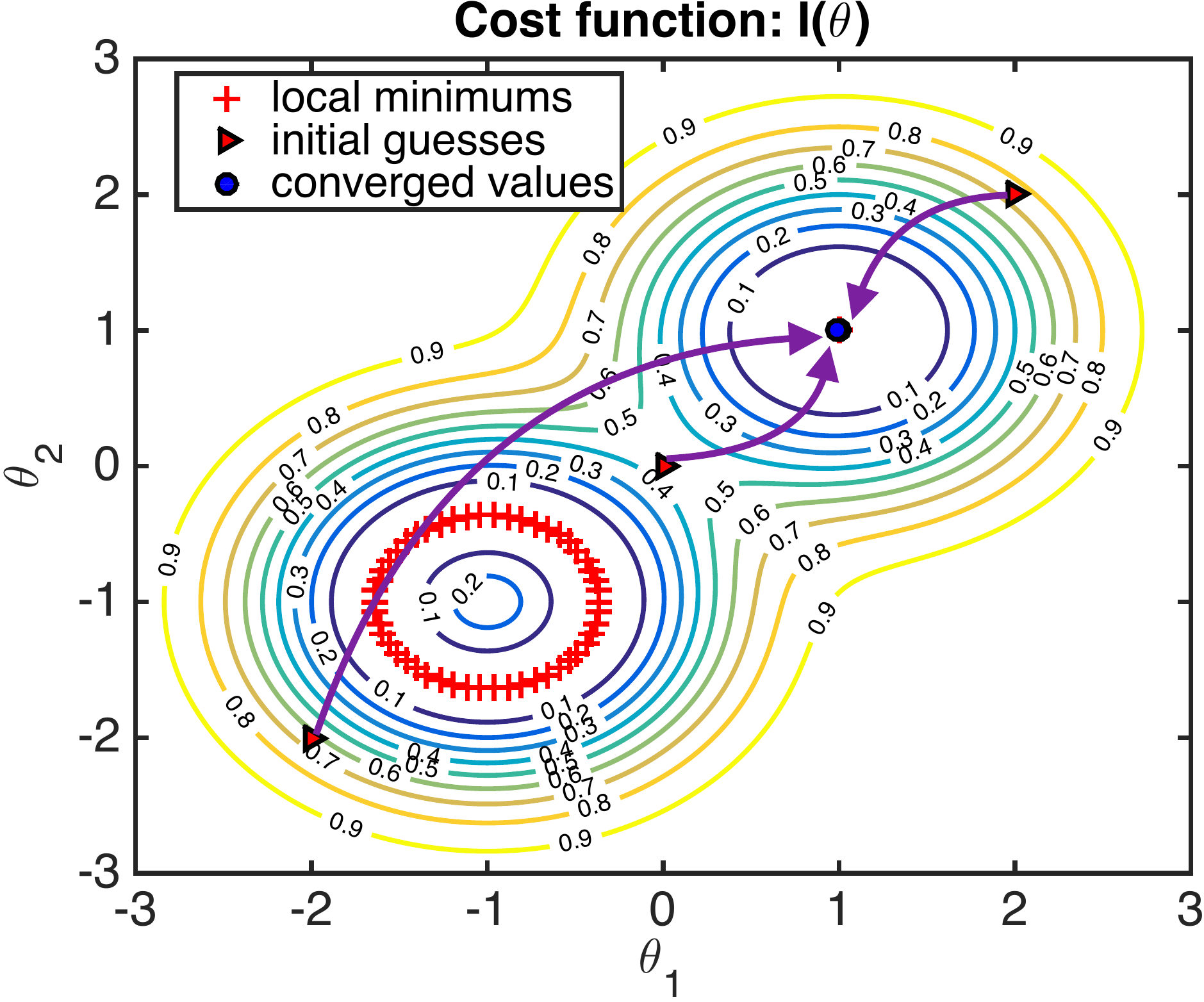}&
    
    \includegraphics[width=0.3\textwidth]{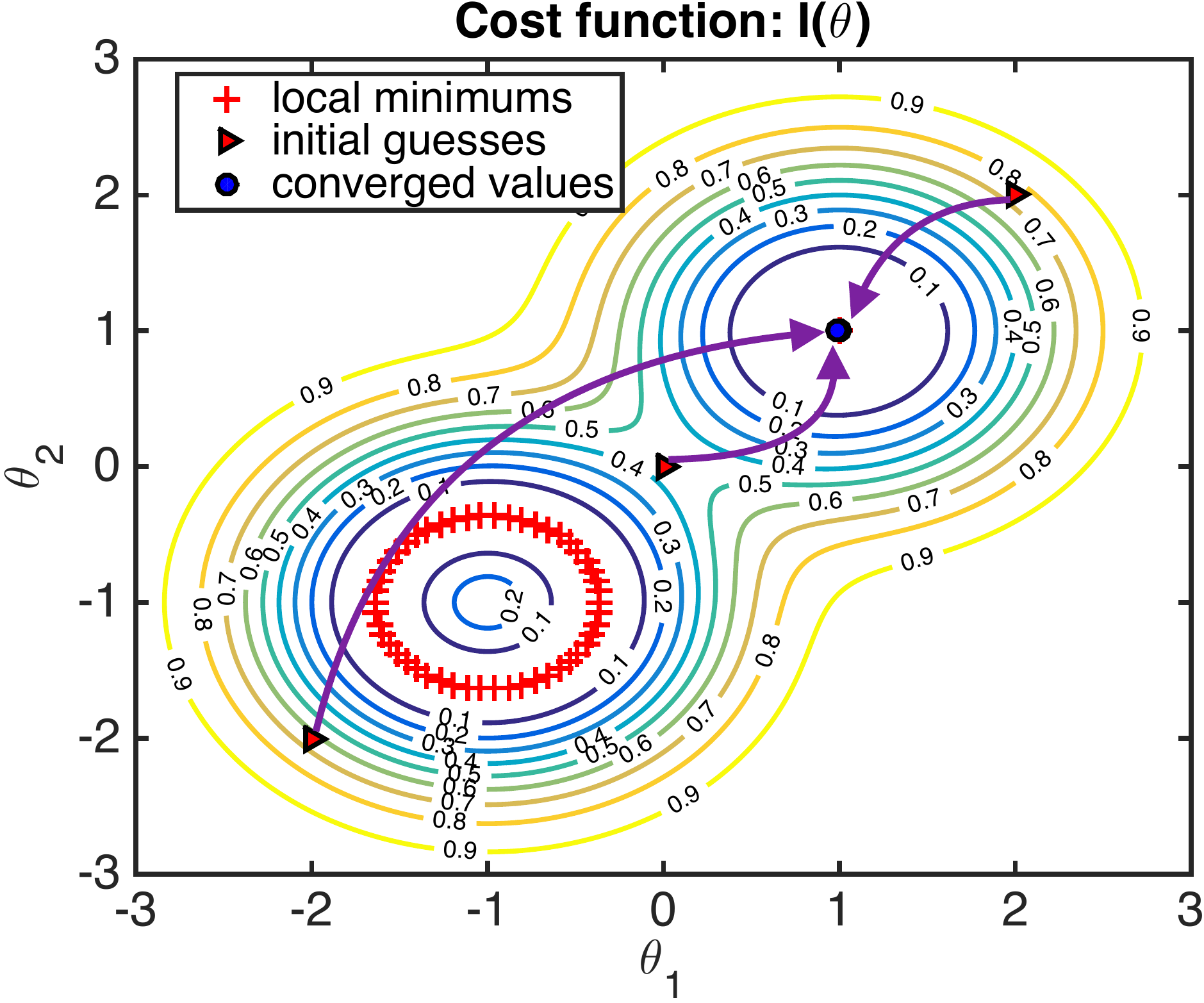}\\
  \end{tabular}
  \caption{Parameter convergence results with constraint imposed. 
  		(Left) When $\Sigma_\theta = [1, 0; 0, 1]$ and $\Sigma_c = 2.0$, the initial guesses at the $(2,2)$ and $(0,0)$ converge to the true local minimum $(1,1)$;
		(Middle) When $\Sigma_\theta = [3, 0; 0, 3]$ and $\Sigma_c = 2.0$, all initial guesses converge to the true local minimum $(1,1)$;
		(Right) When $\Sigma_\theta = [1, 0; 0, 1]$ and $\Sigma_c = 1.0$, all initial guesses converge to the true local minimum $(1,1)$.}
\label{fig:with_constraint_sigma1}
\end{figure*}

Three cases of different initial guesses are simulated with constraint imposed 
by re-weighing individual ensembles based on their likelihood of satisfying the constraint (see \eqref{eq:expectation_with_constraint}).
The covariances are 
$\Sigma_\theta = [1, 0; 0, 1]$ and $\Sigma_l = 0.01$, 
%\begin{equation}
%\Sigma_\theta =
%\begin{bmatrix}
%   1       & 0 \\
%   0       & 1
%\end{bmatrix}
%\;, ~~
%\Sigma_l = 0.01\;,
%\end{equation}
which are kept the same as previous simulations. The covariance of the constraint used here is 
$\Sigma_c = 2.0$,
%\begin{equation}
%\Sigma_c = 2.0\;,
%\end{equation}
which defines a certainty about the constraint.
The simulation results are shown in Table \ref{tab:with_constraint_sigma1} and visualized in Figure \ref{fig:with_constraint_sigma1} (left).
It can be seen that the solution converges to the true value $(1, 1)$ when starting from $(0,0)$ and $(2, 2)$, 
and the solution converges to the local minimum Group II when starting from $(-2,-2)$.
It is interesting to note that the middle initial point $(0,0)$, which originally converges to Group II, now is able to converge to the true value $(1,1)$.

The reason that the solution starting from the lower left initial guess $(-2, -2)$ cannot converges to the true value $(1,1)$ is because 
it is too far way from the true value and the variance of the prior is not large enough to sample the parameter space near the true value $(1,1)$. Therefore, even though the constraint has been imposed, the solution cannot converge to the true value. To verify this, we simulated the three different starting locations with a larger prior variance 
$\Sigma_\theta = [3, 0; 0, 3]$,
%\begin{equation}
%\Sigma_\theta =
%\begin{bmatrix}
%   3       & 0 \\
%   0       & 3
%\end{bmatrix}\;,
%\end{equation}
while the covariance of the data likelihood  $\Sigma_l = 0.01$ and the covariance of the the constraint $\Sigma_c = 2.0$ are kept the same.
The simulation results are shown in Table \ref{tab:with_constraint_sigma3} and visualized in Figure \ref{fig:with_constraint_sigma1} (middle), demonstrating that all the three initial guesses lead to the true parameter value $(1,1)$.

%different initials and ends------------------------------------------------------------
%------------------------------------------------------------

As a further test, we decreased the variance of the constraint $\Sigma_c$ to $1.0$ to see how this influences the parameter estimation.
The results in Table \ref{tab:with_constraint_sigma1_sigma_c1} and Figure \ref{fig:with_constraint_sigma1} (right) demonstrate that (contrary to original conditions in the left panel)
all three initial guesses converge to the true parameter value $\theta^* = (1,1)$. Thus decreasing the variance of the constraint can also improve convergence to the true parameter value in cases where the constraint is more certain.

%\begin{figure}
%    \centering
%    \includegraphics[width=0.5\textwidth]{diff_init_with_constraint_sigma_1}
%    \label{fig:diff_init_with_constraint_sigma_1}
%    \caption{Contour plot of the cost function $I(\theta)$ with respect to parameters $\theta = \left[ \theta_1, \theta_2 \right]$
%    The red ``+" denote the local minimums of the cost function.}
%\end{figure}

\begin{table}[h]%[htbp]
\centering
\scriptsize
\begin{tabularx}{\columnwidth}{cccc}
\toprule
$\theta^0$                   & (-2, -2)                           & (0, 0)                       & (2, 2)\\
\midrule
$\theta^*$                    & (-0.5685, -0.5173)         & (0.9839, 1.0013)    &  (0.9837, 1.0015) \\
\midrule
$HF(\theta^*)$             &  -0.9948                    &-0.9958                        &  -0.9958     \\
\midrule
Group   & II                                  &  I                               & I        \\
\bottomrule
%\label{tb:fully_bayesian}
\end{tabularx}
\caption{Simulation results with the constraint imposed for different initial guesses with $\Sigma_\theta = [1, 0; 0, 1]$, 
    $\Sigma_l = 0.01$ and $\Sigma_c = 2.0$\;.} %\todoJW{$\tau_{\theta}$}}
\label{tab:with_constraint_sigma1}
\end{table}

%%--------------------------
%\begin{figure}
%    \centering
%    \includegraphics[width=0.5\textwidth]{diff_init_with_constraint_sigma_3}
%    \label{fig:diff_init_with_constraint_sigma_3}
%    \caption{Contour plot of the cost function $I(\theta)$ with respect to parameters $\theta = \left[ \theta_1, \theta_2 \right]$
%    The red ``+" denote the local minimums of the cost function.}
%\end{figure}

\begin{table}[h]%[htbp]
\centering
\scriptsize
\begin{tabularx}{\columnwidth}{cccc}
\toprule
$\theta^0$                   & (-2  , -2)                           & (0, 0)                       & (2, 2)\\
\midrule
$\theta^*$                    & (0.9947, 0.9962)        & (0.9958, 0.9943)    &  (0.9955, 0.9942) \\
\midrule
$HF(\theta^*)$             & -0.9859                       &-0.9897                       & -0.9908    \\
\midrule
Group   & I                                  &  I                             & I       \\
\bottomrule
%\label{tb:fully_bayesian}
\end{tabularx}
\caption{Simulation results with the constraint imposed for different initial guesses with $\Sigma_\theta = [3, 0; 0, 3]$, 
    $\Sigma_l = 0.01$ and $\Sigma_c = 2.0$\;.}
\label{tab:with_constraint_sigma3}
\end{table}
%--------------------------

\begin{table}[h]%[htbp]
\centering
\scriptsize
\begin{tabularx}{\columnwidth}{cccc}
\toprule
$\theta^0$                   & (-2, -2)                           & (0, 0)                       & (2, 2)\\
\midrule
$\theta^*$                    & (0.9976, 0.9984)        & (0.9888, 1.0063)    &  (0.9863, 1.0087) \\
\midrule
$HF(\theta^*)$             & -0.9888                       &-0.9957                       & -0.9962    \\
\midrule
Group   & I                                  &  I                             & I       \\
\bottomrule
%\label{tb:fully_bayesian}
\end{tabularx}
\caption{Simulation results with the constraint imposed for different initial guesses with $\Sigma_\theta = [1, 0; 0, 1]$, 
    $\Sigma_l = 0.01$ and $\Sigma_c = 1.0$\;.}
\label{tab:with_constraint_sigma1_sigma_c1}
\end{table}

\subsection{Relation between Bayesian inference and optimization}
Using Bayesian inference framework to estimate the unknown model parameters is intrinsically related to solving a corresponding optimization problem \cite{aravkin2014optimization}.
To see this, we write out the posterior probability distribution of the unknown parameter $\theta$ conditioned on the observed data $\mathcal{D}$ 
and the fact that the constraint needs to be satisfied,
\begin{align}
& p(\theta | \mathcal{D}, G(x) = {\bf 0}) \propto  p(\mathcal{D} | \theta) p(G(x)= {\bf 0} | \theta) p(\theta) \nonumber \\
& = \frac{1}{Z'} \exp\left(-\frac{1}{2} \left\| \Sigma_\theta^{-\frac{1}{2}}(\theta - \hat{\theta}) \right\|^2
-\frac{1}{2} \left\| \Sigma_l^{-\frac{1}{2}}(y-HF(\theta)) \right\|^2 \right. \nonumber \\
& ~~~\left.-\frac{1}{2} \left\| \Sigma_c^{-\frac{1}{2}}  G(F(\theta)) \right\|^2  \right)\;,
\label{eq:posterior_optimization}
\end{align}
where $Z'$ is a normalization constant. 
As mentioned before, the final estimation of the parameter $\theta$ can be taken as the posterior expectation $\mathbb{E}(\theta | \mathcal{D}, G(x) = {\bf 0})$, 
or as the value that maximizes the posterior probability (MAP)
\begin{equation}
\theta^* = \arg\max_{\theta} p(\theta | \mathcal{D}, G(x) = {\bf 0})\;.
\end{equation}
Based on \eqref{eq:posterior_optimization}, solving the MAP estimation of $\theta$ is equivalent to the following  optimization problem:
\begin{equation}
\min_\theta \left\| \Sigma_\theta^{-\frac{1}{2}}(\theta - \hat{\theta}) \right\|^2
+ \left\| \Sigma_l^{-\frac{1}{2}}(y-HF(\theta)) \right\|^2
+ \left\| \Sigma_c^{-\frac{1}{2}}  G(F(\theta)) \right\|^2 \;,
\label{eq:L2}
\end{equation}
where the three terms in the cost function from left to right represent the contributions from the prior, the data and the constraint.
Therefore using Bayesian inference to estimate model parameters is equivalently solving 
an optimization problem of minimizing the miss-match between the observed output and the reconstructed output, 
while penalizing based on the prior and satisfaction of the constraints.
It is easier to see the relations between different terms if we assume all quantities are scalar:
\begin{equation}
\min_{\theta} \frac{1}{\sigma_\theta^2}\|\theta - \hat{\theta} \|^2 + \frac{1}{\sigma_l^2} \| y - HF(\theta) \|^2 + \frac{1}{\sigma_c^2} \| G(F(\theta)) \|^2\;.
\end{equation}
It can be seen that the variances of the prior, the data and the constraints define the relative importance of each individual terms.
The smaller the variance, the more important the corresponding term is in the cost function.
This is reasonable because the information source with smaller belief uncertainty should naturally get more weight.
Increasing the variance of the prior $\sigma_\theta$ not only samples a broader region but also places more relative weight on satisfying the constraint,
which is why increasing the variance of the prior led to convergence of all three initial guesses to the true value 
(see Table \ref{tab:with_constraint_sigma3} and Figure \ref{fig:with_constraint_sigma1} (middle)).
Similarly, decreasing the variance of the constraint can also put more relative weight on the constraint, 
which is verified in the results shown in Table \ref{tab:with_constraint_sigma1_sigma_c1} and Figure \ref{fig:with_constraint_sigma1} (right).
In the cases of multiple constraints, the variance for each constraint can be used to tune the relative importance of the constraint.

\subsection{Extensions}
The constrained Bayesian inference approach here was developed to address non-uniqueness of the solutions for inverse problems. However, it can also be extended as a way to solve more general constrained optimization problems. An advantage of this approach, compared to traditional gradient-based optimization, is that it is derivative-free and does not require construction of the cost function gradient. Gradient information is implicitly represented by the ensemble. This approach can also provide a potential framework to incorporate domain knowledge in learning models to accelerate convergence and improve accuracy.

Although we assumed Gaussian distributions, this approach can be extended to non-Gaussian distributions. This ultimately leads to a different ``weighting" on the ensemble members (see Eq.~(\ref{eq:reweigh})). Applying different distributions for the constraints, prior or data can be useful. For example, assuming a Laplace likelihood for the constraint results in $L_1$-regularization instead of $L_2$-regularization in (\ref{eq:L2}), and a skew likelihood for the constraint will result in different strictness on either side of the constraint surface in parameter space. 

\section{Conclusion}
To address the non-uniqueness of the feasible solutions for ill-posed inverse problems due to model complexity and lack of observation dimension, we here proposed a general method to constrain the inverse problem in a Bayesian inference framework. The constraint is imposed by constructing a likelihood function denoting the fitness of a solution. Then the posterior distribution for the unknown parameter conditioned on both the observation data and the constraint is obtained, and the final parameter estimation is given by the MAP estimation or the posterior mean. This method was also extended to an approximate Bayesian inference framework in terms of the ensemble Kalman filter, which was shown to lead to a re-weighing of ensemble members based on their fitness to the constraint. Numerical simulations were carried out to demonstrate the effectiveness of this approach for basic proof-of-concept. 

\subsection*{Acknowledgements}
This work was supported in part by the American Heart Association, Award No. 18EIA33900046. 

\bibliographystyle{unsrt}
\bibliography{reference}

\end{document}